\documentclass[english,twocolumn,amsmath,amssymb]{revtex4}
\usepackage{amsmath}
\usepackage[usenames]{color}
\usepackage{graphicx}
\usepackage{amssymb}
\providecommand{\tabularnewline}{\\}
\usepackage{graphics}
\usepackage{color}
\usepackage{setspace}
\usepackage{bm}
\DeclareBoldMathCommand{\bnabla}{\nabla}
\usepackage{babel}
\makeatother

\usepackage[normalem]{ulem}  %% Strike through: \sout{}

\usepackage{verbatim}

\begin{document}

\title{Coarse-grained analysis of a lattice Boltzmann model for planar streamer fronts}

%
%	AUTHORS
%
\author{Wim Vanroose}
\altaffiliation[Present address: ]{Departement Wiskunde-Informatica,
Universiteit Antwerpen, Middelheimlaan 1, 2020 Antwerpen, Belgium}
\author{Giovanni Samaey}
\author{Pieter Van Leemput}
\affiliation{Department of Computer Science, Katholieke Universiteit Leuven,
Celestijnenlaan 200A, B-3001 Heverlee, Belgium}
\begin{abstract}
We study the traveling wave solutions of a lattice Boltzmann model for
the planar streamer fronts that appear in the transport of electrons
through a gas in a strong electrical field. To mimic the physical
properties of the impact ionization reaction, we introduce a reaction
matrix containing reaction rates that depend on the electron
velocities. Via a Chapman--Enskog expansion, one is able to find only
a rough approximation for a macroscopic evolution law that describes
the traveling wave solution. We propose to compute these solutions
with the help of a coarse-grained time-stepper, which is an effective
evolution law for the macroscopic fields that only uses appropriately
initialized simulations of the lattice Boltzmann model over short time
intervals. The traveling wave solution is found as a fixed point of
the sequential application of the coarse-grained time-stepper and a
shift-back operator. The fixed point is then computed with a
Newton-Krylov Solver. We compare the resulting solutions with those of
the approximate PDE model, and propose a method to find the minimal
physical wave speed.
\end{abstract}
\maketitle
\date{\today}

%==========================
%INTRODUCTION
%==============================

\section{Introduction}

When a gas of neutral atoms or molecules is exposed to a strong
electrical field, a small initial seed of electrons can
lead to an ionization avalanche. Indeed, the seed electrons are
accelerated by the field and gain  enough energy to ionize the
neutral atoms when they collide.  The two slow electrons that emerge
from this reaction, i.e. the impact and the ionized electron, are
again accelerated by the field and cause, on their turn, an ionization
reaction. Simultaneously the electrical field is locally modified
because of the charge creation. This interplay between the dynamics of
the electrons and the electrical field can lead to a multitude of
phenomena studied in plasma physics such as arcs, glows, sparks and
streamers.

In this article we will focus on the initial field driven ionization
that can lead to traveling waves known as streamer fronts. These waves
have previously been studied by Ebert \emph{et al.} \cite{ebert}
who introduced and analyzed the minimal streamer model, a one-dimensional
model for the propagation of planar streamer fronts. This model consists
of two coupled non-linear PDEs:  a reaction-convection-diffusion
equation for the evolution of the electron density and a Poisson-like 
evolution equation for the electrical field. The reaction term
is based on the Townsend approximation that expresses the growth
of the number of electrons as a function of the local electrical field.

During the last two decades, however, a lot of progress has been made
in the microscopic understanding of impact ionization reactions in
atomic and molecular systems. In this reaction an impact electron
ionizes the target and kicks out an additional electron. There are
several successful theories that can predict the exact probability
distribution of the escaping electron \cite{bray,malegat,pindzola,rescigno}.
 In the next decade,  we expect that the  
theoretical tools will be able to accurately predict the microscopic physics of electron impact on molecular
targets such as N$_{2}$ and O$_{2}$, the most important molecules in
the composition of air.
This progress in the understanding of the impact ionization reaction,
however, has not been incorporated in the description of the
macroscopic behavior such as the minimal streamer front of Ebert
\emph{et al}. Instead, such models still make use of a phenomenological
approximation to the reactions, such as the Townsend
approximation. This article extends the minimal streamer model and incorporates more microscopic
information.  We model the system by 
a Boltzmann equation, which is constructed such that the cross sections in the collision
integral resemble the true microscopic cross sections.

To find the traveling wave solutions of this more microscopic model, we exploit a separation of time scales
between the relaxation of the electron distribution
function to a local equilibrium and the evolution of the macroscopic
fields (electron density and electrical field). It is known from
kinetic theory that the first process is fast: once initialized, it
takes a molecular gas not more than a few collisions to relax to its
equilibrium state.

In kinetic theory, the fast times scales are often eliminated from the
problem by assuming a local equilibrium distribution function which
leads to a reaction-diffusion model with transport coefficients that
depend on the local electrical field. This reduction method, however,
is only successful in the absence of steep gradients in the electron
density \cite{uhlenbeck}, an assumption not valid for the planar
streamer fronts that have, typically, very steep increases in the
electron density.

In this article, we take an alternative route and find the traveling
wave solution through a so-called \textit{coarse-grained time-stepper}
(CGTS) that exploits the separation of time scales to extract 
the effective macroscopic behavior. This method was
proposed by Kevrekidis \emph{et al.} \cite{manifesto} and the numerical aspects 
of its application to find traveling wave solutions of lattice
Boltzmann models have recently been studied \cite{samaey}. The time-stepper 
uses a sequence of computational steps to evolve the
macroscopic state. This sequence involves: (1) a lifting step, which
creates an appropriate electron distribution function for a given
electron density, (2) a simulation step, where the lattice Boltzmann
model is evolved over a coarse-grained time step $\Delta T$, and (3) a restriction step, where the macroscopic
state is extracted from the electron distribution function. This
method does not derive effective equations explicitly, and therefore allows steep gradients to be
present.  We compare our results with those obtained by deriving an approximate macroscopic PDE model through 
the more traditional Chapman-Enskog expansion. 
%%The aim of the paper is to illustrate the applicability of the coarse-grained
The paper therefore illustrates the applicability of the coarse-grained
time-stepper on a non-trivial problem where the exact macroscopic
equations are hard to derive.

The outline of the paper is as follows. In section II, we shortly
review the physics of the impact ionization reaction and the streamer
fronts, and recapitulate the Boltzmann equations.  In section III, we
derive from the Boltzmann equation a lattice Boltzmann model with
multiple velocities and discuss how the ionization reaction, external
field and electron diffusion are incorporated in the model.  Section
IV derives a macroscopic PDE from the model using the Chapman-Enskog
expansion and discusses the minimal velocity of the traveling waves.
Section V formulates the coarse-grained time-stepper and VI how the
traveling wave solutions are found.  Finally in section VII, we have some
numerical results.

%============================
% SECTION MODEL
%========================

\section{Model}

\subsection{The physics of the impact ionization re\label{sub:The-physics-of}action}

The impact ionization reaction is a microscopic reaction where electrons
with, typically, an energy around 50eV collide with an atom or a molecule
and ionize this target. The reaction rates of this process depend
sensitively on a number of parameters. Let us consider the simplest
system: an electron hits a hydrogen atom with a bound electron in
its ground state. When the incoming electron has an energy larger
than the binding energy of the electron in the atom, it can kick out, with a certain probability,
the bound electron that will escape, together with the impact electron, 
from the atom. The total energy of the two electrons after the collision
is equal to the energy of the incoming electron minus the original
binding energy of the bound electron.

The reaction rates of this process are expressed by cross sections;
these are probabilities that a certain event will take place. One
such cross section is the triple differential cross section, which
is the probability to find after the collision one electron
escaping in the direction $(\theta_{1},\phi_{1})$ with an energy
$E_{1}$ and a second electron escaping in the direction $(\theta_{2},\phi_{2})$
with an energy $E_{2}$, where the angles are measured with respect
to the axis defined by the momentum of the incoming electron. Since
the two electrons repel each other, it is more likely that they
escape in opposite directions \cite{wannier}.

\begin{figure}
\resizebox{9cm}{6cm}{\includegraphics{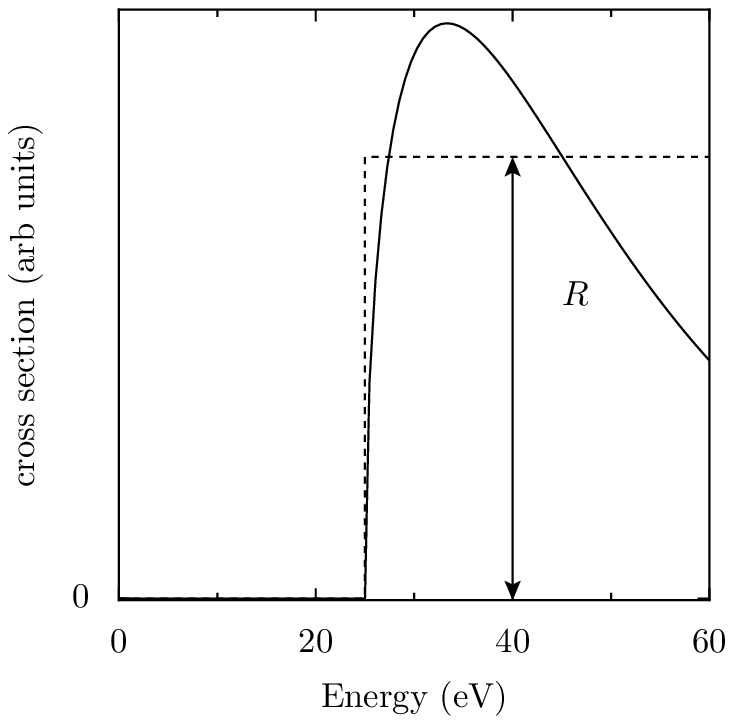}}
\resizebox{9cm}{6cm}{\includegraphics{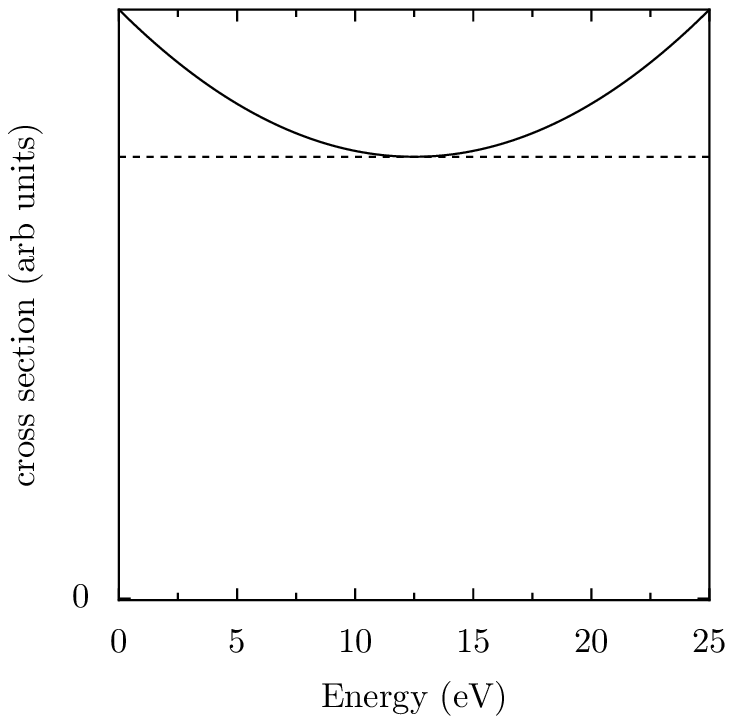}}
\caption{\label{fig:cross_section}
A sketch of the typical shape of the impact ionization cross
section  ( see the experimental results in \cite{hu}). On top, we show the total cross section as a function of the
impact energy where below a threshold energy of 25eV no reactions take
place. In the proposed model we distinguish between slow particles
with an energy below this threshold that do not react and fast
particles with their energy above this threshold. The fast reacting particles
experience a cross section of $R$, as indicated by the dashed line.
In the bottom figure, we show the energy differential cross section
for the escaping electron, where the total energy of the escaping
electrons is 25eV. Since the two electrons are indistinguishable there
is a symmetry.  In our five speed model, we make the approximation
that the two electrons can only escape with equal energy sharing.}
\end{figure}

When this cross section is integrated over all angles
$(\theta_{1},\phi_{1})$ and $(\theta_{2},\phi_{2})$ of the escaping
electrons and all possible ratios of $E_{1}/E_{2}$ of the electron
energies, we get the \emph{total cross section.} This is the total
probability that the incoming electron will cause an ionization
event. This total cross section depends on the energy of the incoming
electron and is zero when the energy of the incoming electron is below
the binding energy of the bound electron. Just above this
binding energy, there is a steep rise in the cross section that is
known as a threshold. Just above this threshold the cross section is
the largest and as we further increase the energy the cross section
diminishes.  This is illustrated in figure \ref{fig:cross_section} (top).

When the cross section is integrated of the angles only, but not over
the relative energies, we get the so called \emph{energy differential
cross section,} which is the probability of causing an ionization
event with a given relative energy of the two electrons. 
In contrast with the electron directions, there is no pronounced preference 
for the energy sharing between the two electrons, see figure \ref{fig:cross_section} (bottom).
It is only slightly more likely that two electrons will come out with unequal energy.

Recently, several theoretical methods have successfully
predicted the directions of the escaping electrons, the total cross
section and the energy differential cross section in the hydrogen
atom. We name \textit{exterior complex scaling} \cite{rescigno},
\textit{time dependent close coupling} \cite{pindzola}, \textit{HRW-SOW} 
\cite{malegat} and \emph{convergent close coupling} \cite{bray}.

When the electron hits a molecular system instead of an atom, the
physics is complicated by the extra degrees of freedom. The cross
sections now depend on both the orientation and internuclear
coordinates of the molecule at the moment of electron impact, as is
seen in processes where two electrons are ejected from molecules after
it is hit by a photon \cite{vanroose,weber}. 
%% Up to now, no exact calculations
%% of the impact ionization cross section in molecular systems have been
%% performed, but we expect this to become possible in the coming years.
Therefore,
%% Because, for molecular systems, the impact ionization reaction rate depends on the orientation
%% and internuclear distances of the molecule at the moment of the electron
%%impact, 
there will be some random terms in the reaction cross section, which will need to be included
in a realistic %%A truly 
microscopic model. 
%%needs to include these random effects. 
In this paper,
we will model the microscopic interactions using a Boltzmann equation, which is still deterministic;
extensions that accurately account for random effects will be treated in future work. 
However, we note that the coarse-grained time-stepper approach that is used in this work has already been applied successfully to study systems with stochastic
effects \cite{qiao}.

\subsection{Review of the physics of streamer fronts.}

Ebert \emph{et al.}~\cite{ebert} introduced the minimal streamer model.
It consists of two coupled non-linear PDEs: a 
reaction-convection-diffusion equation for the evolution of the
electron density and an equation that relates the change in the electrical
field to the charge flux. The electron density evolves because of the drift due to the
electrical field, the electron diffusion and the ionization
reaction, which is formulated in the Townsend approximation.
The reaction rate is then given by an exponential that depends on the strength
of the local electrical field. The evolution of the electrical field
is determined by Poisson's law of electrostatics where the field
changes because of the charge creation by the ionization
reaction. This minimal streamer model exhibits both negatively and
positively charged fronts.  The first moves in the direction of the
electrical field, while the positively charged moves in the opposite
direction and can only propagate because of the electron diffusion and
the ionization reactions. Each of these fronts appears as a one-parameter
family of uniformly translating solutions (since any translate of the wave is also a solution).

In our extension of the Ebert model, we replace the reaction-diffusion
equation for the evolution of the electron density with a Boltzmann
equation for the one-electron distribution function
$f(\mathbf{x},\mathbf{v},t)$, that counts the number of electrons in
the phase-space volume element bounded by position $\mathbf{x}$ and
$\mathbf{x+dx}$ and by speed $\mathbf{v}$ and $\mathbf{v+dv}$. The
Boltzmann equation is 
\begin{equation}
\frac{\partial f(\mathbf{x},\mathbf{v},t)}{\partial t} 
+ \mathbf{v}\cdot\frac{\partial f(\mathbf{x},\mathbf{v},t)}{\partial x} 
+ \mathbf{E}(\mathbf{x},t)\cdot\frac{\partial f(\mathbf{x},\mathbf{v},t)}{\partial v}
= \Omega(\mathbf{x},t) ,
\label{eq:boltzmann}
\end{equation}
where $\mathbf{E}(\mathbf{x},t)$ is the external electrical field and
$\Omega(\mathbf{x},t)$ is the collision operator, an integral operator that
integrates the cross sections of the ionization reaction over the
velocity space.  This Boltzmann equation is coupled to an evolution equation for
the electrical field. Because additional electrons are created, the
local charge density changes the electrical field through the Poisson
law
\[
\bnabla\cdot{\mathbf{E}}(\mathbf{x},t)=q(\mathbf{x},t),\]
 where $q(x,t)=(n_{+}-n_{e})e/q_{0}$ is the charge distribution. Here,
 $n_{+}$ represents the number of ions, $n_{e}$ is the number of electrons and
 $q_{0}$ a unit of charge. 

We now connect the change in electrical field with the change
in the charge distribution. We have
\[
\frac{\partial q(\mathbf{x},t)}{\partial t}+\bnabla\cdot{\mathbf{j}}(\mathbf{x},t)=0\]
in which $\mathbf{j}(\mathbf{x},t)$ is the charge flux.  This leads to an equation for the evolution of the electrical field, 
\begin{equation}
\frac{\partial\mathbf{E}(\mathbf{x},t)}{\partial t}+\mathbf{j}(\mathbf{x},t)=0.\label{eq:electrical_field}\end{equation}
 Since we assume that the ions are immobile, the flux $\mathbf{j}(\mathbf{x},t)$
is solely determined by the one-particle distribution function $f(\mathbf{x},\mathbf{v},t)$
of the electrons.

Our extension of the minimal streamer model is now the coupled evolution
of eq.~(\ref{eq:boltzmann}) and (\ref{eq:electrical_field}).
Note that the set of coupled equations is very similar to the Wigner-Poisson
problem \cite{wigner} used to model electron transport through diodes.

\section{Lattice Boltzmann Discretization}
 Together with the impact ionization cross sections, the coupled
 equations (\ref{eq:boltzmann}) and (\ref{eq:electrical_field}) are a
 non-linear integro-differential equation coupled
 to a scalar partial differential equation for the electrical
 field. This equation in its full dimension is hard to solve, both
 analytically and numerically.  As a first step, we look at the
 one-dimensional streamer fronts of the Boltzmann equation in the
 lattice Boltzmann discretization.

\subsection{Discretization of the Boltzmann equation}
In this section, we discretize the one-dimensional Boltzmann equation
\eqref{eq:boltzmann}. The distribution functions $f(x,v,t)$ are
discretized on a lattice in space, velocity and time. The grid
spacing is $\Delta x$ in space and $\Delta t$ in time. The velocity
grid of $v_{i}$ is chosen such that the distance traveled in a
single time step, $v_{i}\Delta t$, is a multiple of the grid distance
$\Delta x$, or in short 
\[ v_{i}=i\frac{\Delta x}{\Delta
t},\hspace{1cm}\mbox{with}\hspace{1cm}i\in{\mathcal{S}}.  
\]
Typically, only a small set of discrete velocities is used. A
discretization with three grid points on the velocity grid has a set
$\mathcal{S}=\{-1,0,1\}$ and is called a D1Q3 model. A discretization
with five grid points has $\mathcal{S}=\{-2,-1,0,1,2\}$ and is a D1Q5
model. The size of the set $\mathcal{S}$ is denoted by
$m$.   We will also denote $c_i = v_i  \Delta t/ \Delta x$ for the dimensionless velocity.

Note that, for ease of notation, we will also use the set $\mathcal{S}$ to index
matrices.  For example, the result
of a linear operator $A$ working on a vector $v_{j \in \mathcal{S}}$ will
be denoted as $\sum_{i \in \mathcal{S}} A_{ij} v_j$, where both the
indices $i$ and $j$ are in $\mathcal{S}$.  The means that the
$A_{-2,-2}$ matrix element with $\mathcal{S}=\{-2,-1,0,1,2\}$ is the
matrix element in the upper left corner of the matrix.

We start from the continuous equation for the distribution function in
the discrete point $(x+v_{i}\Delta t,v_{i})$ in phase space at time
$t$. In the absence of external forces, the Boltzmann equation in this
point reads
\begin{eqnarray} 
& & \frac{\partial f(x+v_{i}\Delta t,v_{i},t)}{\partial t}+v_{i}\frac{\partial f(x+v_{i}\Delta t,v_{i},t)}{\partial x}\nonumber \\ 
& & =\Omega(x+v_{ i}\Delta t,v_i,t) \label{BtoLBM1}
\end{eqnarray} 
%% PAST NET NIET:
%%\begin{equation} 
%%\frac{\partial f(x+v_{i}\Delta
%%t,v_{i},t)}{\partial t}+v_{i}\frac{\partial f(x+v_{i}\Delta
%%t,v_{i},t)}{\partial x} = \Omega(x+v_{i}\Delta t,t) \label{BtoLBM1}
%%\end{equation} 
A discrete lattice Boltzmann equation is now obtained by replacing the
time derivative with an explicit forward difference, the introduction
of an upwind discretization of the convection term and a downwind
discretization  of the collision term
$\Omega(x+v_{i}\Delta t, v_i,t)$  and replace it with $\Omega(x,v_i,t)$ \cite{cao}, 
\begin{eqnarray}
 &  & \frac{f(x+v_{i}\Delta t,v_{i},t+\Delta t) - 
      f(x+v_{i}\Delta t,v_{i},t)}{\Delta t} \nonumber \\
 &  & + v_{i}\frac{f(x+v_{i}\Delta t,v_{i},t) 
      - f(x,v_{i},t)}{v_{i}\Delta t}=\Omega(x,v_i,t). \label{BtoLBM2}
\end{eqnarray}
Note that this discretization of the spatial derivative becomes
less accurate for the largest speeds in the set $\mathcal{S}$. Indeed,
in the five speed model, for example, the largest speed is $v_{\pm2}=\pm2\Delta x/\Delta t$,
and the convection term will be calculated from the difference between
$f(x+v_{\pm2}\Delta t,v_{\pm2},t)$ and $f(x,v_{\pm2},t)$, which
is $2\Delta x$ apart. The discretization error is then proportional
to $2\Delta x$.

Equation \eqref{BtoLBM2} reduces to
\begin{equation} \label{BtoLBM4}
f_i(x+v_{i}\Delta t,t+\Delta t) - f_i(x,t) = \Delta t \, \Omega_i(x,t),
\end{equation}
where we have introduced the shorthand $f_{i}(x,t)$ for $f(x,v_{i},t)$
and $\Omega_i(x,t)$ for $\Omega(x,v_i,t)$ with $i\in\mathcal{S}$.

\subsection{The collision term}
The collision term consists out of two parts 
\begin{equation} \label{collisionterm}
\Delta t \Omega_{i} = \Omega_{i}^{\mbox{\scriptsize diff}} 
           + \Omega_{i}^{\mbox{\scriptsize reaction}}  
%%           + \Omega_{i}^{\mbox{\scriptsize force}}, 
\end{equation}
where the first term will model the electron diffusion, the second
term the ionization reactions and the third term the influence of the
external force. Note that we also incorporated $\Delta t$ in the
notation. We will now discuss the two terms individually.

The first term models the electron diffusion as a BGK relaxation
process \cite{BGK}.  In this approximation, it assumed that the
distribution is attracted to a local equilibrium distribution function $f_i^{eq}$,
\begin{equation}
\Omega_{i}^{\mbox{\scriptsize diff}}=-\frac{1}{\tau}\left(f_{i}-f_{i}^{eq}\right),
\end{equation}
with $f_{i}^{eq}(x,t)$ the equilibrium distribution for electron
diffusion. In the five speed model, we choose
\begin{equation} \label{BGKeq}
f_{i}^{eq} = w_{i}^{eq} \rho \quad \mbox{with} \quad 
w_{i}^{eq} = \left\{ 0,1/4,2/4,1/4,0 \right\} 
\end{equation}
and $\rho(x,t)$ the electron density.  This choice of equilibrium
weights conserves the number of electrons, but does not conserve
momentum as a traditional fluid would do.  Indeed, the electrons
diffuse because they randomly change their direction during the
elastic collisions with the much heavier neutral molecular
particles. We further chose the weights such that there are no fast
particles under diffusive equilibrium. The relaxation time $\tau$ is
related to the electron diffusion coefficient
\begin{equation}
\tau = \frac{1}{2} + 
\frac{D}{\sum_{i\in\mathcal{S}} c_i^{2} w_{i}^{eq}}
%%\frac{D}{\sum_{i\in\mathcal{S}}v_{i}^{2}w_{i}^{eq}}
\frac{\Delta t}{\Delta x^{2}} \label{tau}. 
\end{equation}
Note that, in the literature, the relaxation time $\tau$ is often characterized by its inverse $\omega = 1/\tau$. 

The second term in \eqref{collisionterm} is the reaction term
$\Omega_{i}^{\mbox{\scriptsize reaction}}$ that is modelled with a
$m \times m$ matrix $\mathcal{R}$ 
\begin{equation}
\Omega_{i}^{\mbox{\scriptsize reaction}} =  \Delta t \sum_{j\in\mathcal{S}} \mathcal{R}_{ij} f_{j},
\end{equation}
which represents the velocity dependent reaction rates and allows us
to select between slow and fast particles.

For the five speed model, we choose a reaction matrix
\begin{equation}
\mathcal{R} = \left(\begin{array}{ccccc} -R & 0 & 0 & 0 & 0\\ R & 0 & 0 & 0 &
R\\ 0 & 0 & 0 & 0 & 0\\ R & 0 & 0 & 0 & R\\ 0 & 0 & 0 & 0 & -R
\end{array}\right) \label{reaction-matrix}
\end{equation}
that describes how the reaction cross sections depend on the
velocities of the particles. Each time step $\Delta t$, a fraction $R$
of the particles with speed $v_{\pm2}$ will collide and cause a
ionization reaction. The reaction rate $R$ is chosen to match the height of the 
cross section, see figure \ref{fig:cross_section}. Since the colliding fast particles transfer their energy
to the bound electron, they will loose energy. Therefore, the number of
particles with speed $v_{\pm2}$ diminishes with a rate $-R\Delta t$
and we have $\mathcal{R}_{-2,-2}=\mathcal{R}_{+2,+2}=-R$. At the same
time, the number of slow electrons increases because both the impact
electron and the ionization electron emerge as slow particles with
speed $v_{\pm1}$. Because of the Coulomb repulsion, the two slow
electrons are more likely to emerge in opposite directions and we
choose the rates such that one electron emerges with speed of $v_{-1}$
and the other with $v_{+1}$.

This choice of model parameters ensures that the energy balance during
the ionization reaction is not violated. As discussed in section
\ref{sub:The-physics-of}, the energy of the incoming electron is
larger that the sum of the energies of the escaping electrons because
some of the impact energy covers the binding energy of the bound
electron. In the above model, a single ionization reaction
transforms one electron with speed $v_{+2}$ into two electrons with,
respectively, speed $v_{+1}$ and speed $v_{-1}$. The energy of the
impact electron is $mv_{+2}^{2}/2 = 2 m \Delta x / \Delta t$, while
the sum of the escaping electrons is merely $2mv_{1}^{2}/2 = m \Delta x
/ \Delta t$, where $m$ is the mass of the electron. So a portion $m
\Delta x/\Delta t$ of the impact energy covers the binding energy.   For our model, this is half of the initial impact energy; for more general problems other values are possible.

%==============================
%BEGIN: THE EXTERNAL FORCE ....
% ==============================

\subsection{External Force \label{external_force}}
We now derive a discretization of the $E \frac{\partial f_i}{\partial v}$
term that models the external force in the Boltzmann equation. We
start by expanding
\[
	\frac{\partial \mathbf{f}}{\partial v} = a_0 \mathbf{v}_0 +
	a_1 \mathbf{v}_1 + \ldots + a_{m-1} \mathbf{v}_{m-1},
\]
where $\mathcal{V} = \{\mathbf{v}_0,\mathbf{v}_1,\ldots,
\mathbf{v}_{m-1} \} $ forms a linear independent set of vectors in
$\mathbb{R}^{m}$. We find the coefficients $a_0,a_1,\ldots, a_{m-1}$
by enforcing the Galerkin condition.
\[
  \left(\frac{\partial \mathbf{f}}{\partial v} - \sum_{i=0}^{m-1} a_i\mathbf{v}_i  \right)\perp \mathcal{V}
\]
In the current paper, we choose a particular set of vectors in
$\mathcal{V}$, namely the polynomials $\{1,v,v^2,\ldots,v^{m-1}\}$
discretized in the points $v_i$. For the five speed example the
vectors are, besides their powers of $\Delta x/\Delta t$,
\[
\mathcal{V} = \left\{ 
\left(\begin{array}{c}1\\1\\1\\1\\1\end{array}\right),
\left(\begin{array}{c}-2\\-1\\0\\1\\2\end{array}\right),
\left(\begin{array}{c}4\\1\\0\\1\\4\end{array}\right),
\left(\begin{array}{c}-8\\-1\\0\\1\\8\end{array}\right),
\left(\begin{array}{c}16\\1\\0\\1\\16\end{array}\right)
 \right\}
\]
The Galerkin condition leads to the linear system 
\begin{equation}
\left(\begin{array}{ccccc}
m & 0 & \alpha & 0 & \beta\\
0 & \alpha & 0 & \beta & 0\\
\alpha & 0 & \beta & 0 & \gamma\\
0 & \beta & 0 & \gamma & 0\\
\beta & 0 & \gamma & 0 & \delta \end{array}\right)\left(\begin{array}{c}
a_{0}\\ a_{1}\\ a_{2}\\ a_{3}\\ a_{4}\end{array}\right) = 
\left(\begin{array}{c} 
\mathbf{v}_0^t \cdot \frac{\partial \mathbf{f}}{\partial v}\\
\mathbf{v}_1^t \cdot \frac{\partial \mathbf{f}}{\partial v}\\
\mathbf{v}_2^t \cdot \frac{\partial \mathbf{f}}{\partial v}\\
\mathbf{v}_3^t \cdot \frac{\partial \mathbf{f}}{\partial v}\\
\mathbf{v}_4^t \cdot \frac{\partial \mathbf{f}}{\partial v}
\end{array}\right) \label{galerkin-condition}\end{equation} 
where $\alpha = \sum_{i\in \mathcal{S}} v_i^2$,$\beta = \sum_{i\in \mathcal{S}} v_i^4$, $\gamma = \sum_{i\in \mathcal{S}} v_i^6$ and $\delta = \sum_{i\in \mathcal{S}} v_i^8$. 

To calculate the right-hand side of \eqref{galerkin-condition} we make a detour around the continuous representation.  We note that  
\begin{eqnarray} 
\mathbf{v}_l^{t} \cdot \frac{\partial \mathbf{f}}{\partial v}  = \int_{-\infty}^{+\infty}v^{l}\frac{\partial f(x,v,t)}{\partial v}dv, 
\end{eqnarray}
where $l\in\{0,1,\ldots,m-1\}$.  Because of our particular choice of
basis vectors and the fact that there are no particles with infinite velocities,
we have that
\begin{eqnarray}
\int_{-\infty}^{+\infty}v^{l}\frac{\partial f(x,v,t)}{\partial v}dv
&& +\int_{-\infty}^{+\infty}f(x,v,t)\frac{\partial
v^{l}}{\partial v}dv\nonumber \\ & &
=v^{l}f(x,v,t)|_{-\infty}^{+\infty}=0\label{partial
integration}\end{eqnarray} 
or, in other words, 
\begin{eqnarray}
\mathbf{v}_l^{t} \cdot \frac{\partial \mathbf{f}}{\partial v} &=& -i\int_{-\infty}^{+\infty}f(x,v,t)v^{l-1}dv \nonumber\\
 &=& -l \sum_{j \in \mathcal{S}} v_j^{l-1} f_j
\label{result_partial_integration}
\end{eqnarray}
With the help of \begin{equation} N
=\left(\begin{array}{ccccc} 0 & 0 & 0 & 0 & 0\\ -1 & -1 & -1 & -1 &
-1\\ -2v_{-2} & -2v_{-1} & -2v_{0} & -2v_{1} & -2v_{2}\\
-3{v_{-2}}^{2} & -3{v_{-1}}^{2} & -3{v_{0}}^{2} & -3v_{{1}}^{2} &
-3{v_{2}}^{2}\\ -4{v_{-2}}^{3} & -4{v_{-1}}^{3} & -4{v_{0}}^{3} &
-4v_{{1}}^{3} & -4{v_{2}}^{3}\end{array}\right),
\end{equation} 
we can now define 
%DEFINITION of  force matrix V_{IJ}
\begin{equation}
V\!\!=\!\!\!\left(\begin{array}{ccccc}
1 & v_{-2} & {v_{-2}}^{2} & {v_{-2}}^{3} & {v_{-2}}^{4}\\
1 & v_{-1} & {v_{-1}}^{2} & {v_{-1}}^{3} & {v_{-1}}^{4}\\
1 & v_{0} & {v_{0}}^{2} & {v_{0}}^{3} & {v_{0}}^{4}\\
1 & v_{1} & {v_{1}}^{2} & {v_{1}}^{3} & {v_{1}}^{4}\\
1 & v_{2} & {v_{2}}^{2} & {v_{2}}^{3} & {v_{2}}^{4}\end{array}\right)\!\left(\begin{array}{ccccc}
m & 0 & \alpha & 0 & \beta\\
0 & \alpha & 0 & \beta & 0\\
\alpha & 0 & \beta & 0 & \gamma\\
0 & \beta & 0 & \gamma & 0\\
\beta & 0 & \gamma & 0 & \delta \end{array}\right)^{-1}\!\!\!\!\!N,\end{equation}
and calculate the external force term as 
\begin{equation} 
E(x,t) \frac{\partial f_{i}(x,t)}{\partial v} = 
E(x,t) \sum_{j\in\mathcal{S}}V_{ij}f_{j}(x,t), \label{external force}
\end{equation} 
where the elements of $\mathcal{S}$ denote matrix elements. 

From eq. (\ref{external force}), it is clear that we can 
include the external force as an additional collision term
in the right-hand side of the lattice Boltzmann equation.

%================================
%  END: THE EXTERNAL FORCE
%=================================

%=====================================
% BEGIN:  THE COLLISIONS TERM
%=====================================

%==============================
%         SECTION FLUX
%==============================

\subsection{Flux}

The evolution of the electrical field $E(x,t)$ is determined by the
net flux $j(x,t)$ of electrons as expressed in
\eqref{eq:electrical_field}.  We discretize
\eqref{eq:electrical_field} on a staggered grid with grid points
halfway between the grid points of the lattice Boltzmann model.  The
flux is defined as the number of particles that move between grid
points (pass through an interface) within a single time step. For the
five speed model, we have 
\begin{eqnarray} 
j(x + \Delta x/2, &t)& 
= f_{1}(x+\Delta x,t) - f_{-1}(x,t) \nonumber \\ 
& & + f_{2}(x+\Delta x,t) - f_{-2}(x,t)\label{flux}\\ 
& & + f_{2}(x+2\Delta x,t) - f_{-2}(x - \Delta x,t) \nonumber 
\end{eqnarray}

\subsection{Coupled equations}
The coupled equations \eqref{eq:boltzmann} and
\eqref{eq:electrical_field} for the evolution of the electron
distribution functions and the electrical field is now, after
discretization,
\begin{multline}
f_{i}(x+v_{i}\Delta t, t+\Delta t) - f_{i}(x,t) = \\
-\frac{1}{\tau} \big( f_{i}(x,t)-f_{i}^{eq}(x,t) \big) 
+\sum_{j\in\mathcal{S}} \Delta t \mathcal{R}_{ij}f_{j}(x,t) \\ 
- \frac{\left(E(x-\Delta x/2,t) + E(x+\Delta x/2 ,t) \right)}{2} 
 \sum_{j \in\mathcal{S}}{\Delta t V_{ij}f_{j}(x,t)}  \label{ionizationLBM} 
\end{multline}
\begin{equation}
E(x  + \frac{\Delta x}{2} ,t+\Delta t ) = E(x  +\frac{\Delta x}{2} ,t) 
- \Delta t j(x + \frac{\Delta x}{2} ,t), \label{ionizationE}
\end{equation}
where $j(x + \Delta x/2 ,t)$ is calculated from \eqref{flux}. 
The equations are
coupled because the electrical field appears as an external force in
the first equation, while the flux drives the evolution of the
electrical field in the second equation. This coupling makes the
evolution of the system non-linear.

\section{A PDE model through Chapman-Enskog expansion \label{sec:Chapman-Enskog-expansion-of}}

The model (\ref{ionizationLBM})--(\ref{ionizationE}) evolves the
electrical field $E(x,t)$ and the distribution functions
$f_{i\in\mathcal{S}}(x,t)$ from $t$ to $t+\Delta t$
simultaneously. Alternatively, the evolution of the distribution
functions can be rewritten in terms of the corresponding
(dimensionless) velocity moments defined as
\begin{equation} \label{momdef}
\varrho_{l}(x,t) = \sum_{i\in\mathcal{S}}c_i^{l} f_{i}(x,t),
\end{equation}
where $l \in \{0,1,\ldots,m-1\}$. 
The zeroth moment $\varrho_{l=0}(x,t)$ corresponds to the electron
density $\rho(x,t)$, i.e.\ the macroscopic variable of interest. The
transformation between distribution functions $f_{i}$ and moments
$\varrho_{l}$ can be written as a matrix transformation $M$. In the
five speed model, this matrix is
\begin{equation} 
M = \left(\begin{array}{rrccc} 1 & 1 & 1 & 1 & 1 \\
-2 & -1 & 0 & 1 & 2\\ 
4 & 1 & 0 & 1 & 4\\ 
-8 & -1 & 0 & 1 & 8\\ 
16 & 1 & 0 & 1 & 16
\end{array}\right) \label{m-matrix}
\end{equation} 
such that $\varrho_{l}=\sum_{i\in\mathcal{S}}M_{li}f_{i}$ and
$f_{i}=\sum_{l=0}^{m-1}{\left(M^{-1}\right)}_{il}\varrho_{l}$.

An evolution law for $\varrho_{l}(x,t)$ is
now easily constructed by the following sequence: first transform
$\varrho_{l}$ into $f_{i}(x,t)$ using $M^{-1}$, then use the lattice
Boltzmann equation \eqref{ionizationLBM} to evolve $f_{i}(x,t)$ to
$f_{i}(x,t+\Delta t)$ and, then, transform back to the moments
$\varrho_{l}(x,t+\Delta t)$.

It has been observed phenomenological that the ionization wave can
approximately be described by a PDE in the density.  This suggests that,
in practice, the evolution of these moments is rapidly attracted to a
low dimensional manifold described by the lowest moment
$\varrho_{0}(x,t)$, which is the density. The higher order moments
have then become functional of this density and the dynamics of the
system can effectively be described by the evolution of this
macroscopic moment.

In general, however, it is very hard to find analytic
expressions for this low dimensional description in the form of a PDE
without making crude approximations. For the problem at hand, we illustrate
these difficulties in this section where we apply the
Chapman-Enskog expansion and derive a macroscopic PDE in terms of electron density. Only after
dropping several coupling terms, a closed PDE is derived.

The model discussed in the previous sections can be summarized by
the lattice Boltzmann equation 
\begin{eqnarray}
 & & f_{i}(x+c_{i}\Delta x,t+\Delta t)-f_{i}(x,t)= \nonumber \\ & &
 -\frac{1}{\tau}\left(f_{i}(x,t)-f_{i}^{eq}(x,t)\right)+\sum_{j\in\mathcal{S}}A_{ij}f_{j}(x,t)\; ,
  \label{model_lbe}
\end{eqnarray} 
for $\forall i\in\mathcal{S}$.  Here, $A_{ij}$ can be the reaction term
$R_{ij}$, a force term $V_{ij}$, or a combination of both.  
%% Tegen nu zullen ze het wel doorhebben zeker? Note again that we use the elements of $\mathcal{S}$ to label the matrix elements. 

%%With the help of a Chapman-Enskog expansion, we will derive an approximate reduced PDE model in terms 
%%of the electron density.

A second order Taylor expansion of the term $f_i(x + c_i \Delta x, t + \Delta t)$ 
in \eqref{model_lbe} around $f_i(x,t)$ leads to
\begin{eqnarray}
 &  & c_{i}\Delta x\frac{\partial f_{i}}{\partial x}+\Delta t\frac{\partial f_{i}}{\partial t}+\frac{c_{i}^{2}\Delta x^{2}}{2}\frac{\partial f_{i}}{\partial x^{2}}\nonumber \\
 &  & +c_{i}\Delta x\Delta t\frac{\partial^{2}f_{i}}{\partial x\partial t}+\frac{\Delta t^{2}}{2}\frac{\partial^{2}f_{i}}{\partial t^{2}}\nonumber \\
 &  & \hspace{0.5cm}=-\frac{1}{\tau}\left(f_{i}-f_{i}^{eq}\right)+\sum_{j\in\mathcal{S}}A_{ij}f_{j},\hspace{1cm}{\forall i\in\mathcal{S}}. \label{CEtaylor}
\end{eqnarray}
 We then expand 
$f_{i}$ in terms of increasingly higher order contributions as follows
\begin{equation} \label{CEfiscaling}
f_{i}=f_{i}^{(0)}+\epsilon f_{i}^{(1)}+\epsilon^{2}f_{i}^{(2)}+\ldots
\end{equation}
with $\epsilon$ a small tracer parameter. In fluid dynamics, $\epsilon$ typically refers to the Knudsen
number. The spatial and time derivatives are scaled respectively as
%%in timescales $t_{0}$, $t_{1}$ and $t_{2}$ 
%%in timescales $t_{0}$, $t_{1}$ and $t_{2}$ 
\begin{equation} \label{CExtscaling}
\frac{\partial}{\partial t}=\frac{\partial}{\partial t_{0}}+\epsilon\frac{\partial}{\partial t_{1}}+\epsilon^{2}\frac{\partial}{\partial t_{2}}
\quad \mbox{and} \quad 
\frac{\partial}{\partial x} = \epsilon \frac{\partial}{\partial x_1},
\end{equation}
where we explicitly presume that a zeroth order time scale $t_0$ is
present in the system. As we will show later on, this scale
corresponds to the observed exponential growth of the electron
density.

Because of the multiple time scales $t_{0}$, $t_{1}$ and $t_{2}$,
all the terms in the expansion will couple to all the time scales,
which complicates the derivation of an effective equation.

In our search for a reduced second order PDE model, we only keep
the terms up to second order in $\epsilon^{2}$. For the same reason,
we also drop the second derivative w.r.t.\ time from \eqref{CEtaylor}.
Substitution of \eqref{CEfiscaling} and \eqref{CExtscaling} into
\eqref{CEtaylor} leads to
\begin{eqnarray}
 & & \epsilon c_{i}\Delta x\frac{\partial f_{i}^{(0)}}{\partial
 x}+\epsilon^{2}c_{i}\Delta x\frac{\partial f_{i}^{(1)}}{\partial
 x}+\epsilon^{2}\frac{c_{i}^{2}\Delta
 x^{2}}{2}\frac{\partial^{2}f_{i}^{(0)}}{\partial x^{2}}\nonumber \\ &
 & +\Delta t\frac{\partial f_{i}^{(0)}}{\partial t_{0}}+\epsilon\Delta
 t\frac{\partial f_{i}^{(1)}}{\partial t_{0}}+\epsilon^{2}\Delta
 t\frac{\partial f_{i}^{(2)}}{\partial t_{0}}\nonumber \\ & &
 +\epsilon\Delta t\frac{\partial f_{i}^{(0)}}{\partial
 t_{1}}+\epsilon^{2}\Delta t\frac{\partial f_{i}^{(1)}}{\partial
 t_{1}}+\epsilon^{2}\Delta t\frac{\partial f_{i}^{(0)}}{\partial
 t_{2}}\nonumber \\ & & +\epsilon c_{i}\Delta x\Delta t 
\frac{\partial^{2}f_{i}^{(0)}}{\partial x\partial
 t_{0}}+\epsilon^{2}c_{i}\Delta x\Delta t
\frac{\partial^{2}f_{i}^{(1)}}{\partial x\partial t_{0}}\nonumber \\
 & & +\epsilon^{2}c_{i}\Delta x \Delta t
 \frac{\partial^{2}f_{i}^{(0)}}{\partial x\partial t_{1}}\nonumber \\
 & = & -\frac{1}{\tau}\left(f_{i}^{(0)}+\epsilon
 f_{i}^{(1)}+\epsilon^{2}f_{i}^{(2)}-f_{i}^{eq}\right)\nonumber \\ & &
 +\sum_{j\in\mathcal{S}}A_{ij}\left(f_{j}^{(0)}+\epsilon
 f_{j}^{(1)}+\epsilon^{2}f_{j}^{(2)}\right)\label{chapman_enskog}
\end{eqnarray}
We will now group the terms order by order and derive expressions for
$f_{i}^{(0)}$, $f_{i}^{(1)}$ and $f_{i}^{(2)}$ and the corresponding
evolution equations for $\rho$ at the different time scales.

We will use the fact that if $\Delta t$ and $\Delta x^2$ are of the
same order of magnitude --- which is the case for our examples --- the
terms that have factors as $\Delta t \Delta x$ are effectively of order
$\Delta x^3$ and can be neglected compared to terms with $\Delta t$ or
$\Delta x^2$.
%============================
%SUBSUBSECTION: ZEROTH ORDER
%============================

%%\subsubsection{Equilibrium distribution function \label{equilibrium_f_0}}
\subsubsection{Zeroth order contribution \label{equilibrium_f_0}}
From expansion \eqref{chapman_enskog}, we collect the zeroth order
terms \begin{equation}
\Delta t\frac{\partial f_{i}^{(0)}}{\partial t_{0}}=-\frac{1}{\tau}\left(f_{i}^{(0)}-f_{i}^{eq}\right)+\sum_{j\in\mathcal{S}}A_{ij}f_{j}^{(0)}.\label{firstorder-expansion}\end{equation}
We choose $f^{(0)}_i$ such that the right-hand side of
\eqref{model_lbe} is zero; these distributions will not evolve on time scale $t_0$ and are found 
as the the solution of the linear system
\begin{equation}
\sum_{j\in\mathcal{S}}\left(1-\tau A_{ij}\right)f_{j}^{(0)}=f_{i}^{eq}.\label{first-order-ls}
\end{equation}
Since $f_{i}^{eq}=w_{i}\rho$ \eqref{BGKeq} only depend on the
density, we find 
$f_{i}^{(0)} = w_{i}^{(0)}\rho$ with the weights
$w_{i}^{(0)}$ defined by
\begin{equation} 
w_{i}^{(0)} = \sum_{j\in\mathcal{S}}\left(\left( 1 - \tau A \right)^{-1}\right)_{ij}w_{j}^{eq}.
\end{equation}

Since the matrix $A$ can include the ionization reaction
that does not conserve the particle number, the sum of the weights
$\sum_{i\in\mathcal{S}}w_{i}^{(0)}$ is not necessarily equal to
one. This means that $\sum_{i\in\mathcal{S}}f_{i}^{(0)}\neq\rho$. We
choose to rescale the weights $w_{i}^{(0)}$ with a normalization
factor $\mathcal{N} = \sum_{i\in \mathcal{S}} w_{i}^{(0)} =
\sum_{i,j\in\mathcal{S}} \left(\left(1-\tau A\right)^{-1}\right)_{ij} w_{j}^{eq}$ such
that $\sum_{i\in\mathcal{S}}f_{i}^{(0)}=\rho$.  With this rescaling we
find the zeroth order term of the Chapman-Enskog expansion
\begin{equation}
f_{i}^{(0)}=w_{i}^{(0)}\rho=\frac{1}{\mathcal{N}}\sum_{j\in\mathcal{S}}\left(\left(1-\tau
A\right)^{-1}\right)_{ij} w_{j}^{eq}\rho\label{ce_zeroth_order}
\end{equation}

The rescaling, however, forces us to reconsider equation
\eqref{first-order-ls} because $f_{i}^{(0)}$, as defined above, fails
to be a solution. Still, we keep our $f_{i}^{(0)}$ of
\eqref{ce_zeroth_order} as our zeroth-order term and find for the
evolution of the system at time scale $t_0$
\begin{eqnarray}
 &  & \Delta t\frac{\partial f_{i}^{(0)}}{\partial t_{0}}=-\frac{1}{\tau}\sum_{j\in\mathcal{S}}\left(1-\tau A_{ij}\right)f_{j}^{(0)}+\frac{1}{\tau}f_{i}^{eq}\nonumber \\
 &  & \hspace{0.5cm}=-\frac{1}{\tau}\sum_{j\in\mathcal{S}}\left(1-\tau A_{ij}\right)\sum_{k\in\mathcal{S}}\frac{1}{\mathcal{N}}\left(\left(1-\tau A\right)^{-1} \right)_{jk} f_{k}^{eq}+\frac{1}{\tau}f_{i}^{eq}\nonumber \\
 &  & \hspace{0.5cm}=\frac{1}{\tau}\left(1-\frac{1}{\mathcal{N}}\right)f_{i}^{eq}\nonumber \\
 &  & \hspace{0.5cm}=\alpha f_{i}^{eq}  \Delta t , \label{eq:evolution_f_0}
\end{eqnarray}
 where the growth factor is 
\begin{equation}
\alpha=(1-1/\mathcal{N})/(\tau \Delta t) . \label{eq:growth_factor}
\end{equation}

Summation of \eqref{eq:evolution_f_0} over the set $\mathcal{S}$ leads
to the zeroth order PDE for the evolution of $\rho$ 
\begin{equation}
\frac{\partial\rho}{\partial t_{0}}=\alpha\rho.
\end{equation}
This is a growth equation if $\alpha$ is positive, which is the case
for the ionization reaction.

%=========================
%subsection FIRST ORDER
%=========================

\subsubsection{First order contribution}

To derive the first order equation, we collect the terms that are
first order in $\epsilon$ from \eqref{chapman_enskog}
\begin{eqnarray}
c_{i}\Delta x\frac{\partial f_{i}^{(0)}}{\partial x}+\Delta t\frac{\partial f_{i}^{(1)}}{\partial t_{0}}\nonumber \\
+\Delta t\frac{\partial f_{i}^{(0)}}{\partial t_{1}}+c_{i}\Delta t\Delta x\frac{\partial^{2}f_{i}^{(0)}}{\partial x\partial t_{0}}\nonumber \\
=-\frac{1}{\tau}\sum_{j\in\mathcal{S}}\left(1-\tau A_{ij}\right)f_{j}^{(1)}. \label{first-order-expansion}
\end{eqnarray}
We drop the term $c_{i}\Delta t\Delta
x\partial^{2}f_{i}^{(0)}/(\partial x\partial t_{0})$ because it is of
order $\Delta t\Delta x$, which is smaller than the other terms that
are first order in $\Delta t$ or $\Delta x$. The second term we
neglect is $\Delta t\partial f_{i}^{(1)}/\partial t_{0}$ because we will show below, a postiori, that  it is also of order $\Delta t \Delta x$.

We now have 
\[
c_{i}\Delta x\frac{\partial f_{i}^{(0)}}{\partial x}+\Delta t\frac{\partial f_{i}^{(0)}}{\partial t_{1}}=\sum_{j\in\mathcal{S}}\left(-\frac{1}{\tau}+A_{ij}\right)f_{j}^{(1)}\]
what leads to a first order term 
\begin{equation}
f_{i}^{(1)}=\sum_{j\in\mathcal{S}}\left( \left(-1/\tau+A\right)^{-1} \right)_{ij} \left(c_{j}\Delta x\frac{\partial f_{j}^{(0)}}{\partial x}+\Delta t\frac{\partial f_{j}^{(0)}}{\partial t_{1}}\right). \label{eq:first_order}
\end{equation}
We now see that it is justified to neglect the term $\Delta t \partial f_{i}^{(1)}/ \partial t_0$ in \eqref{first-order-expansion} because it is of order $\Delta t \Delta x$.
%%Using the result $f_{i}^{(0)}=w_{i}^{(0)}\rho$ from the previous
%%section and that from
%%$\sum_{i\in\mathcal{S}}f_{i}=\sum_{i\in\mathcal{S}}\left(f_{i}^{(0)}+\epsilon
%%f_{i}^{(1)}+\epsilon^{2}f_{i}^{(2)}\right)=\rho$ and
%%$\sum_{i\in\mathcal{S}}f_{i}^{(0)}=\rho$ follows that the sum
%%$\sum_{i\in\mathcal{S}}f_{i}^{(1)}=0$,
Using \eqref{ce_zeroth_order} and
$\sum_{i\in\mathcal{S}}f_{i}^{(1)}=0$ (the latter because 
$\sum_{i\in\mathcal{S}}f_{i}=\sum_{i\in\mathcal{S}}\left(f_{i}^{(0)}+\epsilon
f_{i}^{(1)}+\epsilon^{2}f_{i}^{(2)}\right)=\rho$ and
$\sum_{i\in\mathcal{S}}f_{i}^{(0)}=\rho$), 
we find the following PDE at time scale $t_1$ for the evolution of the system
\begin{equation}
\frac{\partial\rho}{\partial t_{1}}+\mathcal{C}\frac{\partial\rho}{\partial x}=0,\label{equation_first_order}
\end{equation}
where the advection coefficient $c$ equals
 \begin{equation}
\mathcal{C}=\frac{\sum_{i,j\in\mathcal{S}}\left(\left(-1/\tau+A\right)^{-1} \right)_{ij}c_{j}w_{j}^{(0)}}{\sum_{i,j\in\mathcal{S}}\left(\left(-1/\tau+A\right)^{-1} \right)_{ij} w_{j}^{(0)}}\frac{\Delta x}{\Delta t}.
\label{eq:c}
\end{equation}
With the help of \eqref{equation_first_order}, the first order
contribution (\ref{eq:first_order}) can be written alternatively as
\begin{equation}
f_{i}^{(1)}=\sum_{j\in\mathcal{S}} \left(\left(-1/\tau+A\right)^{-1} \right)_{ij}w^{(0)}(c_{j}\Delta x-\mathcal{C}\Delta t)\frac{\partial\rho}{\partial x}.\label{first-order-rewrite}
\end{equation}

%=========================
%subsection SECOND ORDER
%=========================

\subsubsection{Second order contribution}
Finally, we derive the second order evolution. We collect
from \eqref{chapman_enskog} the second order terms and find 
\begin{eqnarray}
c_{i}\Delta x\frac{\partial f_{i}^{(1)}}{\partial x}
+ \frac{c_{i}^{2}\Delta x^{2}}{2}\frac{\partial^{2}f_{i}^{(0)}}{\partial x^{2}}+ \Delta t\frac{\partial f_{i}^{(2)}}{\partial t_{0}} \nonumber \\
+ \Delta t\frac{\partial f_{i}^{(1)}}{\partial t_{1}}+ \Delta t\frac{\partial f_{i}^{(0)}}{\partial t_{2}} + c_{i}\Delta t\Delta x\frac{\partial^{2}f_{i}^{(1)}}{\partial x \partial t_{0}}\nonumber \\
+ c_{i}\Delta t\Delta x\frac{\partial^{2}f_{i}^{(0)}}{\partial x\partial t_{1}} = -\frac{1}{\tau}\sum_{j\in\mathcal{S}}\left(1-\tau A_{ij}\right)f_{j}^{(2)}
\end{eqnarray}
The terms $\partial^{2} f_{i}^{(1)}/\left(\partial x \partial t_{0}
\right)$, $\partial^{2} f_{i}^{(0)}/\left(\partial x \partial t_{1}
\right)$, and $\partial f_i^{(1)}/\partial t_1$ (when replacing $f^{(1)}_i$ by 
\eqref{eq:first_order}) are of order $\Delta t \Delta
x$, which is smaller than $\Delta x^2$ for our parameter settings.  
We also neglect $\Delta t \partial f_i^{(2)}/\partial t_0$ because it
can be shown, again a postiori, that it is of order $\Delta t \Delta x$.

%%After these approximations, t
The second order expansion term then becomes
\begin{eqnarray}
c_{i}\Delta x\frac{\partial f_{i}^{(1)}}{\partial x}+\frac{c_{i}^{2}\Delta x^{2}}{2}\frac{\partial^{2}f_{i}^{(0)}}{\partial x^{2}}\nonumber \\
+\Delta t\frac{\partial f_{i}^{(0)}}{\partial t_{2}}=-\frac{1}{\tau}\sum_{j\in\mathcal{S}}\left(1-\tau A_{ij}\right)f_{j}^{(2)},
\end{eqnarray}
%% We insert the expression \eqref{first-order-rewrite} for 
%%in the first term and replace $f_i^{(0)}$ with \eqref{ce_zeroth_order}. We find 
When replacing $f_{i}^{(1)}$ and $f_i^{(0)}$ by \eqref{first-order-rewrite} and \eqref{ce_zeroth_order}, we get 
\begin{eqnarray}
c_{i}\Delta x\sum_{j\in\mathcal{S}} \left(\left(-1/\tau+A\right)^{-1} \right)_{ij}w_{j}^{(0)}\left(c_{j}\Delta x
- \mathcal{C}\Delta t\right)\frac{\partial^{2}\rho}{\partial x^{2}}\nonumber \\
+ \frac{c_{i}^{2}\Delta x^{2}w_{i}^{(0)}}{2}\frac{\partial^{2}\rho}{\partial x^{2}}
+ \Delta t w_i^{(0)} \frac{\partial \rho}{\partial t_{2}}\nonumber \\
= - \frac{1}{\tau}\sum_{j\in\mathcal{S}}\left(1-\tau A_{ij}\right)f_{j}^{(2)}
\end{eqnarray}
and the expression for the second order term becomes
\begin{eqnarray}
f_{i}^{(2)} &=&\sum_{j\in\mathcal{S}}B_{ij}c_{j}\sum_{j\in\mathcal{S}}B_{jk}\left(c_{k}\Delta x-\mathcal{C} \Delta t\right)\Delta  x w_{k}^{(0)}\frac{\partial^{2}\rho}{\partial x^{2}} \nonumber \\
&&+\sum_{j\in\mathcal{S}}B_{ij}c_{j}^{2}w_{j}^{(0)}\frac{\Delta x^{2}}{2}\frac{\partial^{2}\rho}{\partial x^{2}} \nonumber \\
&& +\sum_{j \in\mathcal{S}}  B_{ij}w_{j}^{(0)}\Delta t\frac{\partial\rho}{\partial t_{2}},\end{eqnarray}
where we use $B_{ij}=\left((-1/\tau+A)^{-1}\right)_{ij}$. 
If we define 
\begin{eqnarray}
\mathcal{D}&=& \left(\sum_{i,j,k\in\mathcal{S}}B_{ij}c_{j} B_{jk}\left(c_{k}\Delta x-\mathcal{C}\Delta t\right)\Delta xw_{k}^{(0)}  \right.  \label{eq:diffusion} \\
&& + \left.  \sum_{k,i\in\mathcal{S}}B_{ij}c_{j}^{2}w_{j}^{(0)}\Delta x^2/2 \right)
/ \left( -\sum_{i,j\in\mathcal{S}}B_{ij}w_{j}^{(0)}  \Delta t \right),  \nonumber 
\end{eqnarray}
we obtain the evolution at time scale $t_{2}$ (because $\sum f_{k}^{(2)}=0$)
\begin{equation}
\frac{\partial\rho}{\partial t_{2}}=\mathcal{D}\frac{\partial^{2}\rho}{\partial^{2}x},
\end{equation}
where $\mathcal{D}$ acts as a diffusion coefficient.

\begin{figure}
\resizebox{9cm}{5cm}{\includegraphics{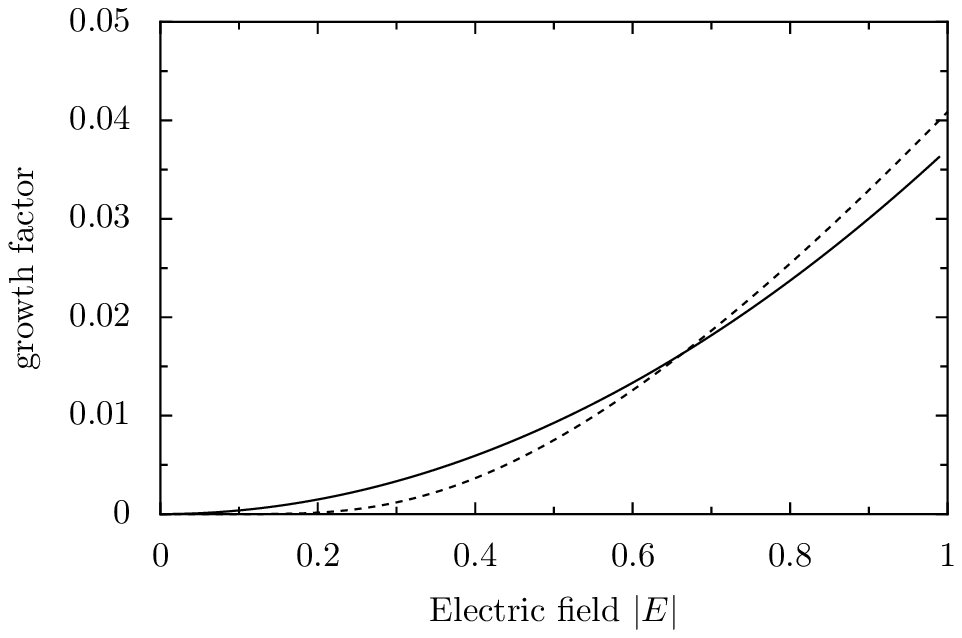}}
\resizebox{9cm}{5cm}{\includegraphics{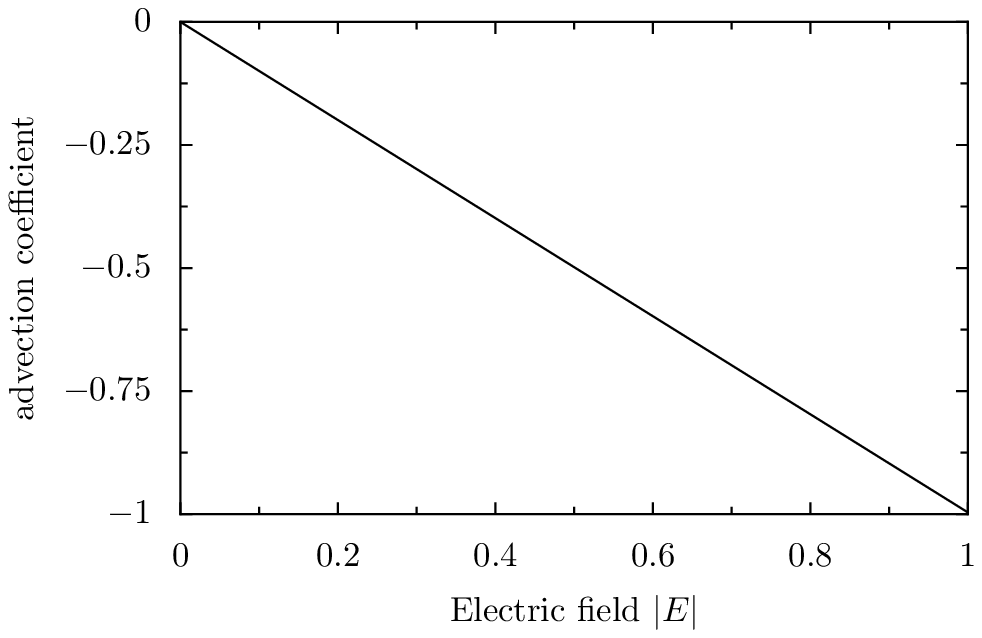}}
\resizebox{9cm}{5cm}{\includegraphics{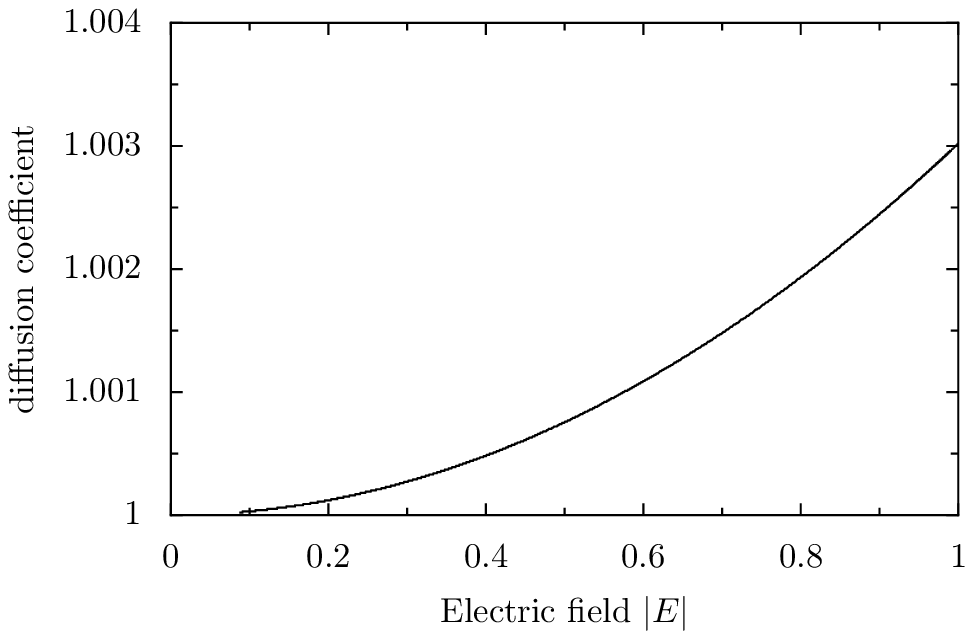}}
\caption{
Top: The growth factor $\alpha(E)$ (solid) of the PDE model derived from
the lattice Boltzmann model. The growth depends on the strength of the
local electrical field and is similar to the Townsend approximation
with $0.111\cdot |E|\exp(-1/|E|)$(dashed).  We have a
reaction rate $R=100$ and model parameters given in section
\ref{sec:Numerical-Results}. Middle: The advection coefficient $\mathcal{C}$ is equal to the external field $-E$.
Bottom: The diffusion coefficient $\mathcal{D}$ only changes with the external field  in the fourth significant figure.
\label{growth_factor}}
\end{figure}

We are now in the position to combine the evolution at the different
timescales $t_{0}$, $t_{1}$ and $t_{2}$ into a single PDE. Because the
matrix $A$ of the model equation \eqref{model_lbe} contains both the
reaction term and the external force, the transport coefficients $\alpha$, $\mathcal{C}$
and $\mathcal{D}$ will depend on the local electrical field $E(x,t)$.
For our example the dependence of the transport coefficients is shown in figure \ref{growth_factor}, we find that $\mathcal{D}$
hardly depends on the strength of the field and can be set equal to
the electron diffusion $D$ used in \eqref{tau} to define the relaxation of
the lattice Boltzmann model.  In a similar way, we find that $c$, the
transport coefficient of the advection term, is approximately equal to
the $-E$, the local electrical field that causes the drift.  Only the
growth factor $\alpha$, defined in \eqref{eq:growth_factor}, depends
on the strength of the local electrical field.  With the help of these
observations we get the coupled PDE
\begin{eqnarray}
\frac{\partial\rho}{\partial t} & = & \alpha ( E(x,t)) \rho 
+ E(x,t)\frac{\partial\rho}{\partial x}
+ D\frac{\partial^{2}\rho}{\partial x^{2}} \nonumber \\
\frac{\partial E}{\partial t} & = & - E(x,t)\rho
- D \frac{\partial\rho}{\partial t} \label{coupled_system_CE}
\end{eqnarray}
for the evolution of $E(x,t)$ and $ \rho(x,t)$. The second equation expresses the flux with the help of the transport coefficients.

These coupled equations are similar to minimal streamer model of
Ebert, Van Saarloos and Caroli \cite{ebert}, except that the growth
rate is now defined by \eqref{eq:growth_factor} instead of the
Townsend approximation. In figure \ref{growth_factor}, we illustrate how the growth
coefficient depends on the local electrical field and compare with a
Townsend approximation. We find that a Townsend reaction term $ 0.111\cdot |E|\exp(-1/|E|)$ approximately describes  a similar
growth term as the PDE model derived from the lattice Boltzmann model.

\subsubsection{Traveling wave solutions \label{sub:traveling-wave-solution}}
The system \eqref{coupled_system_CE} is non-linear and it is
well-known that it has a one-parameter family of front solutions that
translate uniformly with a speed $c$
\cite{wvansaarloos}. 
There is a minimal speed $c^{*}$ that is usually found by looking at
the asymptotic region $\rightarrow+\infty$.  In this limit, the
electrical field becomes constant and is denoted by $E^{+}$ --- the
same notation as in \cite{ebert} --- and the equation for the electron
density becomes 
\begin{equation} \label{PDEasy}
\frac{\partial\rho}{\partial t}=\alpha(E^{+})\rho+E^{+}\frac{\partial\rho}{\partial x}+D\frac{\partial^{2}\rho}{\partial x^{2}},
\end{equation}
where the transport coefficients have become constants. In a co-moving
coordinate frame that travels with the same speed $c$ along the
x-axis, we define $\xi=x-ct$. The PDE \eqref{PDEasy} becomes
stationary and the solution in the asymptotic region fits
\begin{equation}
0=\alpha(E^{+})\rho(\xi)+(c+E^{+})\frac{\partial\rho(\xi)}{\partial\xi}+D\frac{\partial^{2}\rho(\xi)}{\partial\xi^{2}}.
\end{equation}
The latter is a second order ODE that can be transformed into two
coupled first order ODEs by denoting $\partial \rho / \partial \xi$ as
$v$ and $\rho$ as $u$ . The system of coupled equations is 
%%\begin{eqnarray}
%%u_{\xi} & = & v\\ v_{\xi} & = &
%%-\frac{\alpha(E^{+})}{D}u-\frac{c+E^{+}}{D}v
%%\end{eqnarray} 
%%We rewrite this in matrix notation 
%%\pvlcomment{Plaatsbesparing: vergelijkingen hierboven zijn exact dezelfde als 
%%die in matrixvorm hieronder}
\begin{equation}
\left(\begin{array}{c}
u_{\xi}\\ v_{\xi}\end{array}\right)=\left(\begin{array}{cc} 0 & 1\\
-\frac{\alpha(E^{+})}{D} &
-\frac{c+E^{+}}{D}\end{array}\right)\left(\begin{array}{c} u\\
v\end{array}\right)
\end{equation}
where $u_\xi$ and $v_\xi$ denote derivatives of, respectively, $u$ and
$v$ to $\xi$. The matrix has two eigenvalues
\[
\lambda_{\pm}=\frac{-(c+E^{+})\pm\sqrt{(c+E^{+})^{2}-4D\alpha(E^{+})}}{2D}.
\]
There are two cases, if $(c+E^{+})^{2}<4D\alpha(E^{+})$ the two
eigenvalues are complex, otherwise, they are real.

The electron density in the asymptotic region is now a linear combination
of two exponentials 
\begin{equation} \label{combexp}
\lim_{x\rightarrow+\infty}\rho(x)=Ae^{\lambda_{+}x}+Be^{\lambda_{-}x}.
\end{equation}
When the two eigenvalues are complex, the asymptotic density is
oscillating and can becomes negative. This is unphysical because we
cannot have a negative number of particles and it is concluded that
the speed $c$ has to be above a minimal speed 
\begin{equation}
c>c^{*}=c(E^{+})+2\sqrt{ D\alpha(E^{+})}, \label{minimal_speed}
\end{equation} 
to keep both eigenvalues real. Note that both eigenvalues
$\lambda_{+}$ and $\lambda_{-}$ coalesce at the critical speed
$c=c^{*}.$

\section{The coarse-grained time-stepper \label{sub:Coarse-Grained-Time}}

\begin{figure*}
\begin{tabular}{ccc}
%	MICROSCOPIC BOX
 
%   WAVE graphics
\resizebox{6cm}{3cm}{\includegraphics{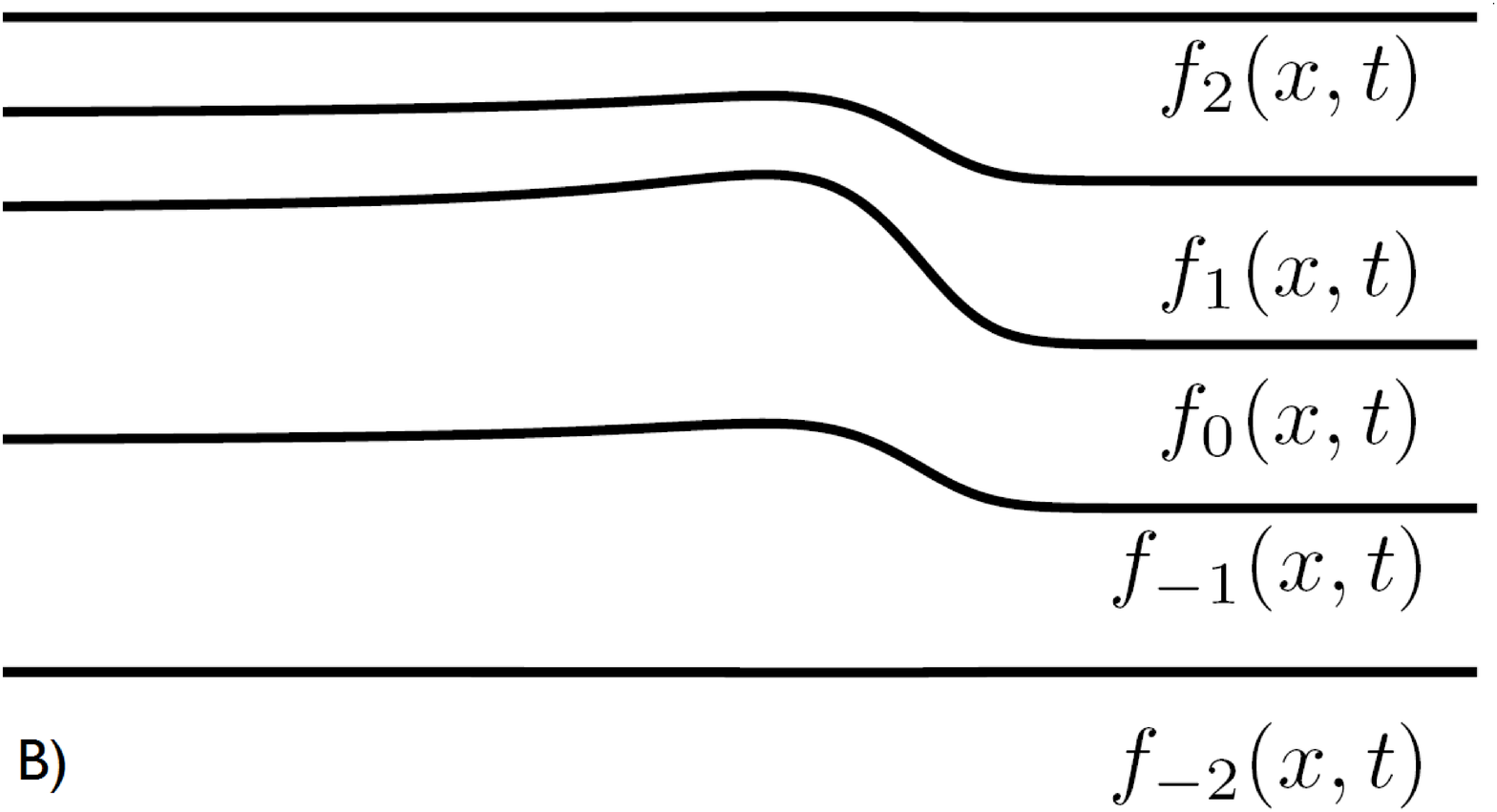}}&\fbox{\rotatebox{90}{\hspace{0.4cm}$\downarrow$\hspace{0.4cm}LBM\hspace{0.4cm}$\downarrow$\hspace{0.4cm} }}&\resizebox{6cm}{3cm}{\includegraphics{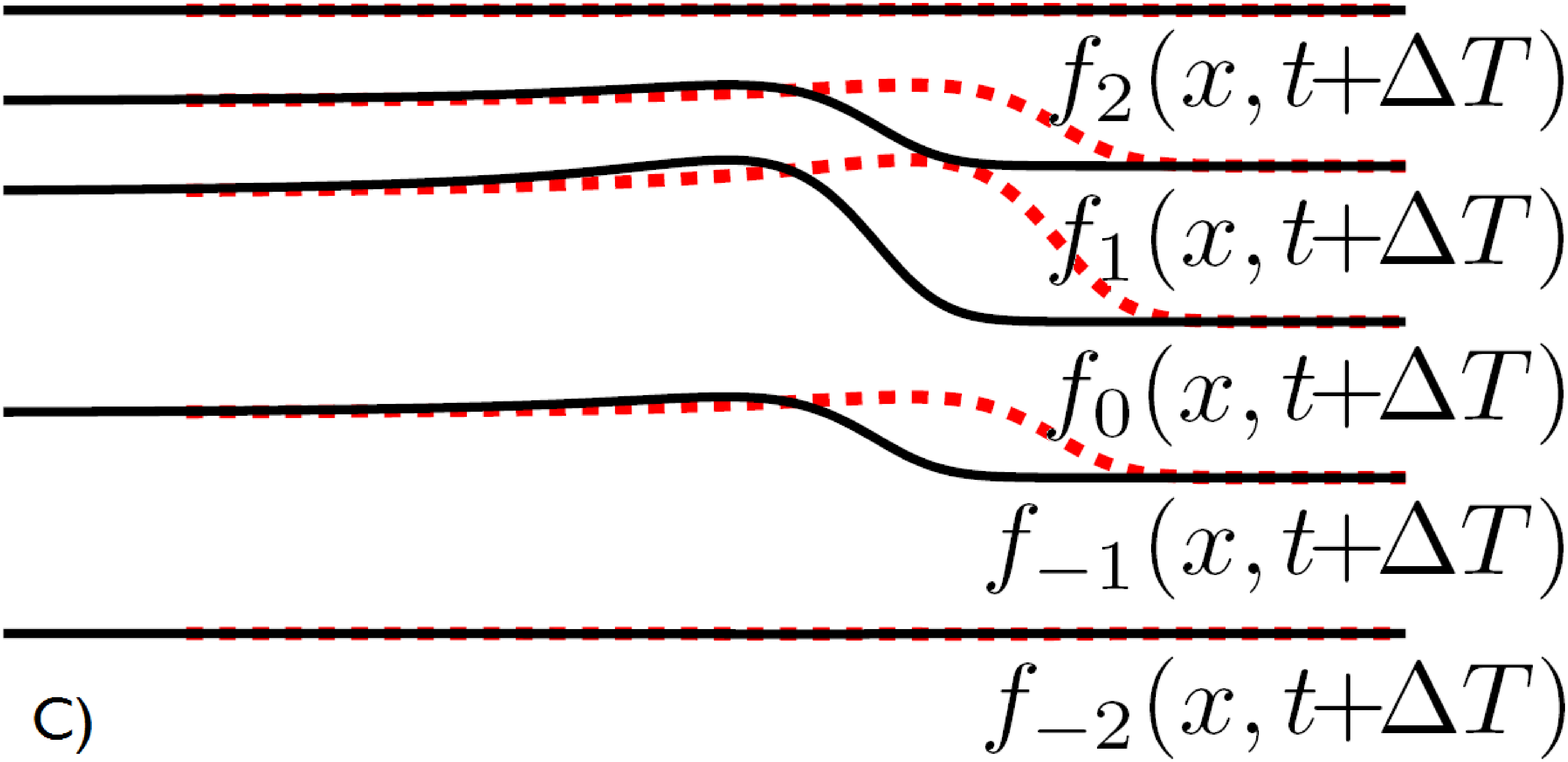}}\\
 \fbox{ $\uparrow$ LIFT $\uparrow$} &&  \fbox{ $\downarrow$ RESTRICT $\downarrow$}\\
 \resizebox{6cm}{3cm}{\includegraphics{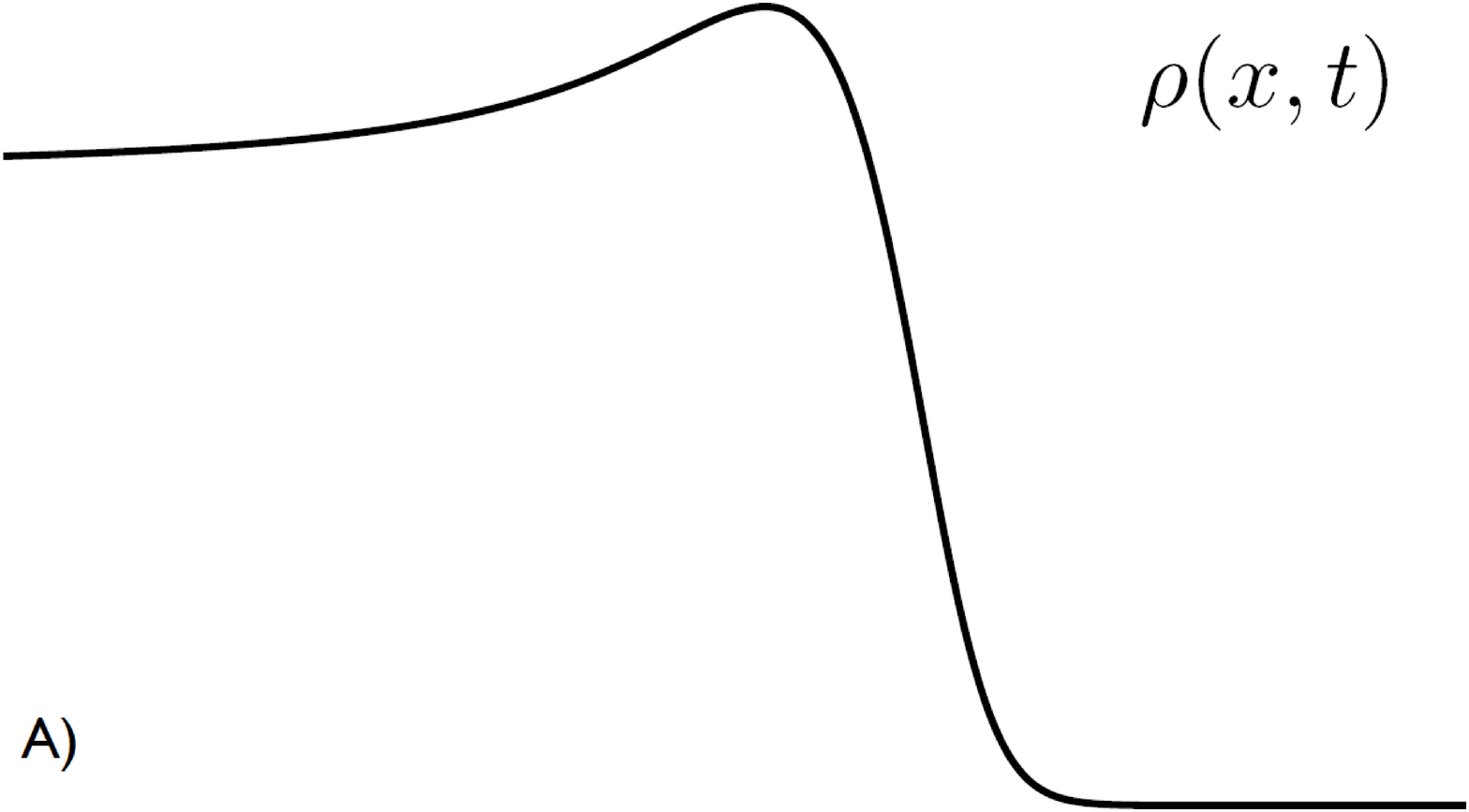}}&\fbox{\rotatebox{90}{\hspace{0.3cm}$\uparrow$\hspace{0.3cm}SHIFT-BACK\hspace{0.3cm}$\uparrow$\hspace{0.3cm}}}&\resizebox{6cm}{3cm}{\includegraphics{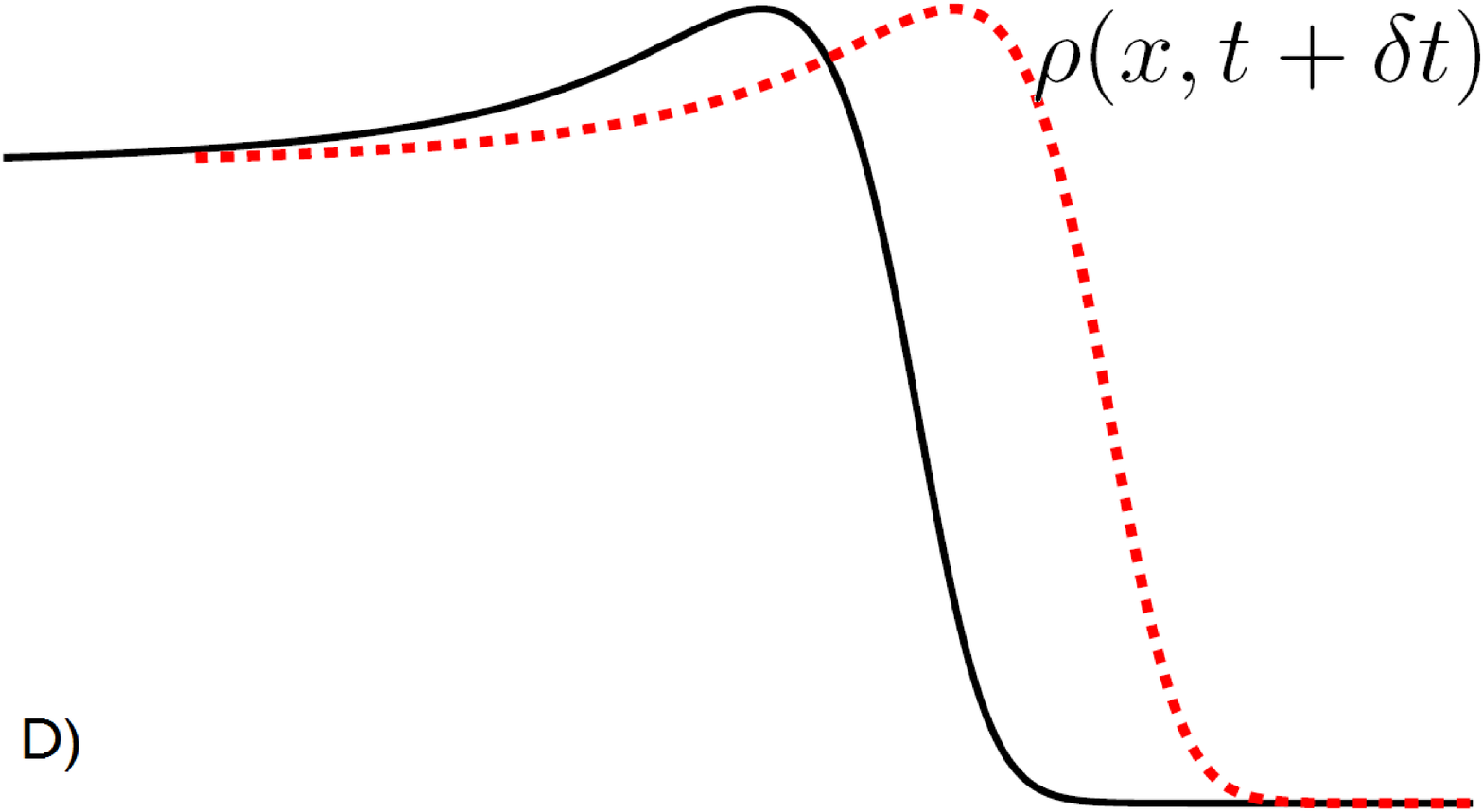}}
\end{tabular}
\caption{\label{fig:micmac}
A summary of the different steps involved in finding the traveling
wave solutions as fixed points of the coarse-grained time-stepper. We
start with an initial guess for the density in the left bottom (panel
a). This density is mapped to the corresponding components of the
one-particle distribution function using the constrained runs lifting operator. The
initial conditions (panel b) are then evolved with the full lattice
Boltzmann model over a time $\Delta T$. Each component travels over a
distance $c\Delta T$ and arrives at panel c. In the next step, the
density at time $t + \Delta T$ is extracted using the restriction
operator (panel d). The resulting density is shifted back over a
distance $c \Delta T$ to arrive at the original position. The
traveling wave solution should be invariant under this sequence of
operations and is formulated as a fixed point.}
\end{figure*}

In this section, we describe an alternative way to perform the analysis of
the macroscopic behavior of the system. It is based on the work of Kevrekidis
\emph{et al.} \cite{manifesto} who developed a coarse-grained time
stepper (CGTS), which is an effective evolution law for the
density. This evolution law $\mathcal{F}$ is not an analytic
expression such as a PDE, but the following sequence of computational
steps: 1) \textit{lifting}, 2) \textit{simulation} and 3)
\textit{restriction}, denoted by the operators $\mu$, LBM and 
$\mathcal{M}$ respectively (figure \ref{fig:micmac}).   Note that the simulation time 
$\Delta T$ is in general a multiple of $\Delta t$, the lattice Boltzmann time step.  
Formally, this is written as
\begin{eqnarray}
U(x,t+\Delta T) & = & \mathcal{F}(U(x,t), \Delta T)\nonumber \\
 & = & \mathcal{M}(\mbox{LBM}(\mu(U(x,t)),\Delta T)),\label{eq:cgts}
\end{eqnarray}
where we have introduced $U(x,t)=(\rho(x,t),E(x,t))$ as a shorthand notation.
The time-stepper $\mathcal{F}$ evolves the macroscopic density
$\rho(x,t)\!=\!\sum_{i\in\mathcal{S}}f_{i}(x,t)$ and the electrical field
$E(x,t)$ from time $t$ to
$t+\Delta T$. 

%%\begin{equation}
%%\rho(x,t+\Delta T)=\mathcal{F}(\rho(x,t)),
%%\end{equation}
%%
%%The coarse-grained time stepper is then defined as the sequence of
%%lifting, simulation and restriction \
%%
%%It is clear that the lifting step is the most critical step in the
%%proposed sequence. Indeed, there are multiple choices for $f_{j}(x_{i},t)$
%%that lead to the same $\rho(x,t)$ and the details of this lifting
%%are discussed in the next section.
%%
%%
%%The lifting maps the density to the corresponding distribution
%%functions, the simulation evolves these distribution functions, and,
%%finally, the restriction extracts the density at $t+\Delta T$. This
%%sequence constructs an effective evolution law for the density by
%%passing through the microscopic description, our case the mesoscopic
%%lattice Boltzmann model.
%%
%% We will discuss the details of this lifting step further down this section.
%%\subsection{Lifting with constrained runs}

\subsection{Lifting} \label{sec:lifting}

Since the electrical field $E(x,t)$ is the same in both the lattice Boltzmann and the macroscopic model, we can ignore it for the discussion of the lifting and restriction operators.
In the \textit{lifting} step, the particle distribution functions are
initialized starting from the initial density
\begin{equation*}
\mu:\mathbb{R}^{n}\mapsto\mathbb{R}^{n\times m}:\rho(x_{j},t)\mapsto f_{i}(x_{j},t)
\end{equation*}
with $i\in\mathcal{S}$, $m$ the number of speeds in $\mathcal{S}$, and  $j\in\{1,2,\ldots,n\}$ denoting the discrete spatial 
grid points. Because lifting is a one-to-many mapping problem, it is
the most critical step in the coarse-grained time-stepper.  We use the
\textit{constrained runs scheme}, an algorithm proposed in \cite{gear}
in the context of singularly perturbed systems. 
Here, it is wrapped around a single time step $\Delta t$ of the
lattice Boltzmann model \cite{pvl}. 

The procedure is given in table \ref{crl}.
\begin{table}
\hrule \vspace{2pt}
 \textbf{Algorithm 1} Constrained runs scheme for LBM \vspace{2pt}
 \hrule \vspace{2pt}
 \begin{tabular}{cll}
\textbf{initialize} &
$f_{i}^{[0]}=w_i^{eq}\rho(x,t)$ &
$\forall i\in\mathcal{S}$\tabularnewline
\textbf{repeat} &
&
\tabularnewline
&
$f^{[k+1]}=\mbox{LBM}(f^{[k]})$ &
a single LBM step\tabularnewline
&
$\varrho^{[k+1]}=Mf^{[k+1]}$ &
map into moments\tabularnewline
&
$\rho^{[k+1]}=\rho(x,t)$ &
reset the density\tabularnewline
&
$f^{[k+1]}=M^{-1}\varrho^{[k+1]}$ &
map into distributions \tabularnewline
\textbf{until} &
convergence heuristic \vspace{2pt}
 &
\tabularnewline
\end{tabular}\hrule
\caption{
Lifting. The constrained runs algorithm computes the distribution functions
$f_{i}(x,t)$ corresponding to a given density $\rho(x,t)$. The superscript $k$ indicates the iteration number.
\label{crl}}
\end{table}
Starting from an initial guess $\rho(x,t)$ for the density,
we obtain initial guesses for the distribution functions using the BGK equilibrium weights \eqref{BGKeq}.  This choice determines
the initial guess for the higher order moments through the transformation
matrix $M$, see equation \eqref{m-matrix}.  (In principle, the initial guesses for the higher order moments can be chosen arbitrarily; the scheme is designed precisely to converge to the correct value of these moments for the given density.) 

We then perform the following iteration. First, we use the 
lattice Boltzmann model to evolve
$f_{i\in\mathcal{S}}^{[k]}$ from $t$ to $t+\Delta t$. The result is
transformed back into the moment representation by a matrix
multiplication with $M$, which gives us $\varrho^{[k+1]}$. Next, the
zeroth moment of the vector $\varrho^{[k+1]}$ is
reset to the initial value $\rho(x,t)$. Transforming this modified
moment vector $\varrho^{[k+1]}$ back into distribution functions gives
us the next $f_{i\in\mathcal{S}}^{[k+1]}$.  We repeat this iteration until the
higher order moments have converged.

\begin{figure}
\begin{center}
\resizebox{9cm}{5cm}{\includegraphics{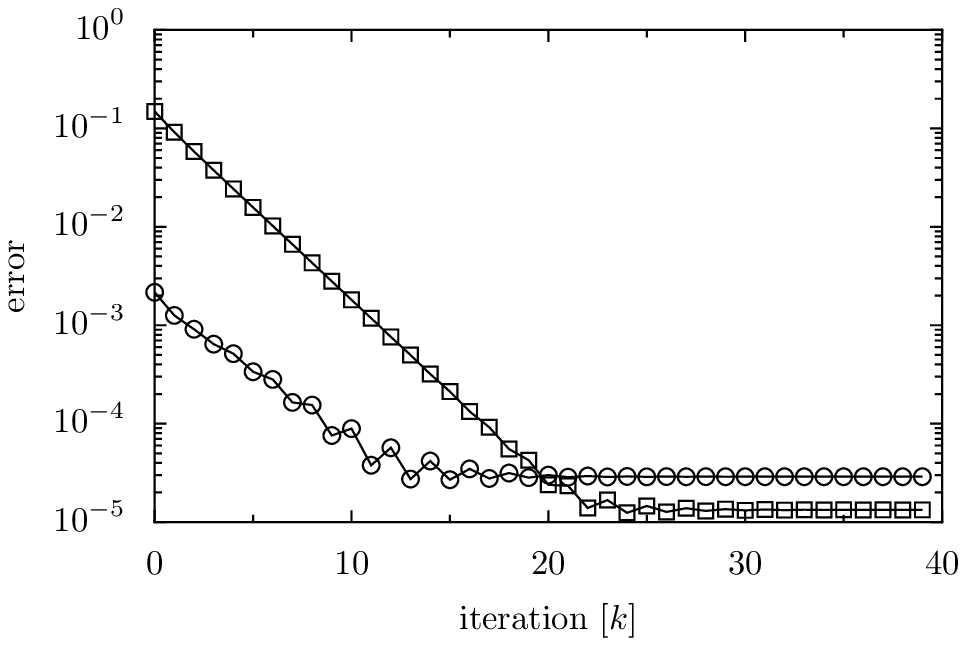}}\\
\resizebox{9cm}{5cm}{\includegraphics{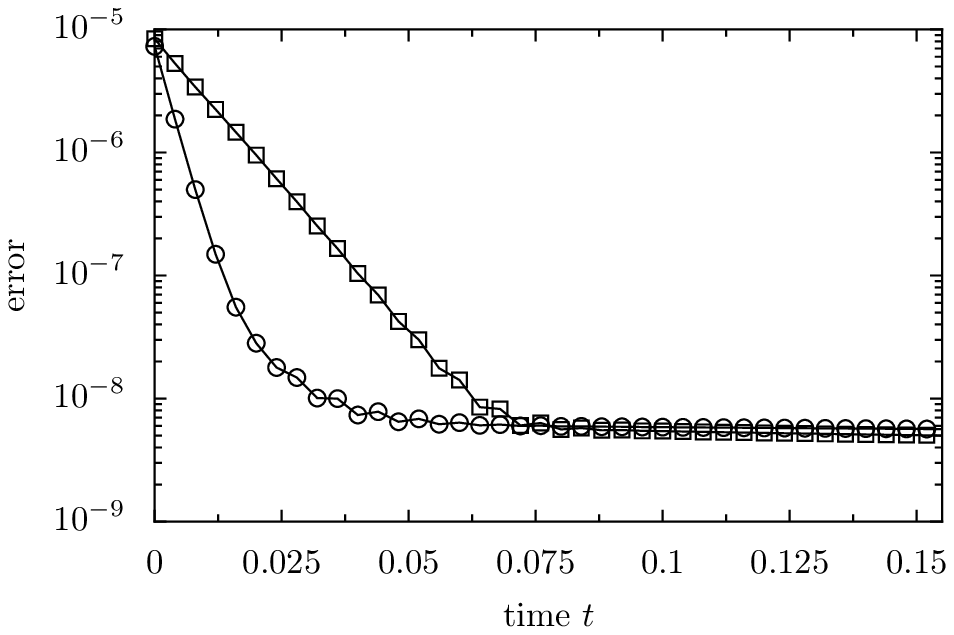}}\\
\caption{
Convergence of the constrained runs algorithm for the five speed
ionization model both during lifting and simulation step. Top:
Error (2-norm) in the lifted distribution functions 
(circles) and the flux (squares) as a function of the number
constrained runs iterations.  After an initial convergence with rate 
$|1 -1/\tau - \Delta t R|$, the error stagnates after approximately 25
iterations.  Bottom: Difference (2-norm) between the distribution functions
(circles) and the flux (squares) of the lattice Boltzmann simulation 
that started from the stagnation point of
the top figure and the original simulation. Again the error decreases
for approximately 25 lattice Boltzmann steps.  Note however
that after stagnation there is a slow evolution because the macroscopic fields also evolve.\label{figure_crl}}
\end{center}
\end{figure}

The convergence behavior of the constrained runs algorithm from table
\ref{crl} is analyzed in \cite{pvl} for one-dimensional
reaction-diffusion lattice Boltzmann models with $\mathcal{S} =
\{-1,0,1\}$ (D1Q3 stencil) and a density dependent reaction term.
For such systems, the algorithm is unconditionally stable and
converges up to the first order terms in the Chapman-Enskog expansion
of the distribution functions.  The convergence rate is $|1-1/\tau|$,
i.e.\ the same rate at which the lattice Boltzmann simulation
relaxes towards the diffusive BGK equilibrium.

Below, we extend the results from \cite{pvl} to the current five speed
model with the velocity dependent reaction term
\eqref{reaction-matrix}. In the absence of an electrical force field,
the distributions $f_{\pm2}$ for the fast particles evolve as
\eqref{ionizationLBM}, i.e.
\begin{eqnarray}
f_{\pm2}(x_j \pm 2 \Delta x,t+\Delta t) &=& 
 (1-\frac{1}{\tau} - \Delta t R) f_{\pm2}(x_j,t) \nonumber \\
 &+& \frac{1}{\tau} w^{eq}_{\pm2} \rho(x_j,t), \label{LBMsimulf2}
\end{eqnarray}
where the second term is ``frozen'' because in each iteration of the
constrained runs algorithm, the density is reset to
its original value. Because the LBM propagation of distributions is a
conservative operation \cite{succi}, this iteration is linearly stable
if
\begin{equation} \label{5speedconvrate}
|1 - \frac{1}{\tau} - \Delta t R| < 1. 
\end{equation}

%% PVL: (note that, specifically, we have $w^{eq}_{\pm2} = 0$ here).

%%Omdat $w_{\pm2}^{eq} = 0$ sterft $f_{\pm2}$ uit zonder electrical field.

The distributions $f_{\pm1}$ evolve as
\begin{eqnarray}
&&f_{\pm1}(x_j \pm \Delta x,t+\Delta t) = 
 (1-\frac{1}{\tau}) f_{\pm1}(x_j,t) \nonumber \\
 &+& \frac{1}{\tau} w^{eq}_{\pm1} \rho(x_j,t) 
 + \Delta t R \big( f_{+2}(x_j,t) + f_{-2}(x_j,t) \big), 
\label{LBMsimulf1}
\end{eqnarray}
where the number of slow particles is increased proportional to the
number of fast particles because of the ionization reaction.  Again,
the density value in the BGK equilibrium in \eqref{LBMsimulf1} is
``frozen'' to the initial value.  Because the convergence rate for the
$f_{\pm2}$ components is given by \eqref{5speedconvrate}, equation
\eqref{LBMsimulf1} converges at a rate $|1-1/\tau - \Delta t R|$ if
this value dominates over $|1 - 1/\tau|$, or at a rate $|1-1/\tau|$ if
the latter is dominant over \eqref{5speedconvrate}.

%We conclude that the constrained runs scheme becomes unstable when
%\eqref{5speedconvrate} does not hold, i.e.\ when the underlying
%lattice Boltzmann simulation is unstable.

%%{\color{red} but we expect that the results hold
%%because the external force can be incorporated within the equilibrium
%%distribution and that does not alter the convergence properties.}  
%%\pvl{Ik zou dit weglaten: het is me niet zo duidelijk, immers BGK 
%%evenwicht hangt af van $\rho$, terwijl electrical force term
%%rechtstreeks op $f_j$'s werkt.} 

So far, we have no formal proof for the convergence when an electrical
field is present, but we can illustrate the convergence of the
algorithm for the full system
(\ref{ionizationLBM})--(\ref{ionizationE}) numerically for the
parameter settings from section~\ref{sec:Numerical-Results}. The
figure is produced as follows. We first extract the velocity moments
\eqref{momdef} from a full lattice Boltzmann simulation that has
evolved from an initial state for several thousand time steps. Subsequently,
we use the obtained density $\rho$ as the initial condition for
another lattice Boltzmann simulation and use algorithm \ref{crl} for
its initialization; the distribution functions of the original simulation are considered to be the ``exact'' solution. 
Figure \ref{figure_crl} (top) plots the norm of the
error between the constrained runs state and the state of the first
lattice Boltzmann simulation that it tries to reconstruct. We also observe
that \emph{after} initialization, when evolving the full lattice 
Boltzmann system both from the ``exact'' and the re-initialized 
distributions, the 
error between the first and second system further decreases (figure \ref{figure_crl}, bottom). The same
observation was made in \cite{samaey}.  Note that we have no
analytical expression for the initial state returned by the
constrained runs scheme in this setting, but we believe that the
results on the accuracy from \cite{pvl} generalize and that the
obtained initial state is a first order approximation of the
Chapman-Enskog relations.

\subsection{Simulation}

In the \textit{simulation} step, the initial distributions obtained
from the lifting step, are evolved for a coarse-grained
macroscopic time step $\Delta T$ using the lattice Boltzmann model
discussed in the previous sections. This step is denoted as
\[
f_{i}(x_{j},t+\Delta T)=\mbox{LBM}(f_{i}(x_{j},t),\Delta T),
\]
where the number of time steps depends on the ratio of $\Delta
T/\Delta t$.  When the ratio is not an integer, a linear interpolation
is used between subsequent steps.

\subsection{Restriction}

In the last step, the \textit{restriction} step, we extract the
macroscopic variables from the result of the simulation. The
macroscopic density at time $t+\Delta T$ is then
\begin{multline}
\mathcal{M} : \mathbb{R}^{n\times m}\mapsto\mathbb{R}^{n} 
: f_{i}(x_{j},t+\Delta T)\mapsto \\\rho(x_{j},t+\Delta T)=\sum_{i\in\mathcal{S}}f_{i}(x_{j},t+\Delta T).  
\end{multline}

%==================================================
% SECTIE over FIXED POINT
%==================================================

\section{The traveling waves as a fixed point problem}

In this section we describe the methodology outlined in
\cite{samaey} to find the traveling wave solutions of the
coarse-grained time-stepper $\mathcal{F}(U(x,t))$ defined in section
\ref{sub:Coarse-Grained-Time}. 
If a traveling wave solution with a speed $c$ of $\mathcal{F}(U(x,t))$
is evolved over a time $\Delta T$, the solution has shifted
over a distance $c\Delta T$. We define a shift-back operator
$\sigma_{\psi}$ that shifts the solution back over a distance $\psi.$
\[
\sigma_{\psi}:U(x,t)\mapsto\sigma_{\psi}(U(x,t))=U(x,t)+\psi\partial_{x}U(x,t),
\]
 where we implement the shift-back by using the characteristic solution
of $\partial_{t}U(x,t)+\psi\partial_{x}U(x,t)$ in the forward Euler
time discretization.

This shift-back operator is combined with the coarse-grained time-stepper 
to write a non-linear system for the traveling wave in the co-moving coordinate system with $\xi = x -c t$
\begin{equation} \label{fixedpointproblem}
U(\xi)-\sigma_{c\Delta T}\left(\mathcal{F}(U(\xi),\Delta T)\right)=0.
\end{equation}
This equation expresses the sequence of computational steps as illustrated in figure \ref{fig:micmac}. 
This system, however, is
singular because any translate of a solution will also be a solution
of \eqref{fixedpointproblem} \cite{samaey}.

To get a regular system we add phase (pinning) condition $p(U)$ and 
a regularization parameter $\alpha$
as an additional unknown, as discussed in \cite{samaey}.  The resulting
non-linear system is
\begin{equation}
\mathcal{G}(U,\alpha)=\left\{ \begin{array}{lcc}
U-\sigma_{c\Delta T}\left(\mathcal{F}(U,\Delta T)\right) & = & 0\\
p(U) & = & 0\end{array}\right.\label{eq:non-linear system},\end{equation}
where the phase condition $p(U)$ is defined as
\[
p(U)=\int_{\xi_{0}}^{\xi_{N-1}}U(\xi)d_{\xi}U_{ref}(\xi)d\xi.
\]
This condition minimizes phase shifts with respect to the reference solution  $U_{ref}(\xi)$.

\subsection{Preconditioned Newton-GMRES}

We solve the non-linear system (\ref{eq:non-linear system}) using
a Newton-Raphson method, 
\[
\left\{ \begin{array}{lcr}
U^{[k+1]} &=&U^{[k]}+dU^{[k]}\\
\alpha^{[k+1]} &=&\alpha^{[k]}+d\alpha^{[k]}\end{array},\right.\]
 where the corrections $dU^{[k]}$ and $d\alpha^{[k]}$ are calculated
by solving, in each Newton iteration, a linear system of the form
\begin{eqnarray}
&& \left(\begin{array}{cc}
I-J(U^{[k]},\alpha^{[k]}) & d_{\xi}U^{[k]}\\
d_{U}p(U^{[k]}) & 0\end{array}\right)\left(\begin{array}{c}
dU^{[k]}\\
d\alpha^{[k]}\end{array}\right)\nonumber\\
 &&  =  -\mathcal{G}(U^{[k]},\alpha^{[k]}).\label{eq:linear_system}
\end{eqnarray}
This system is the linearization of $G(U,\alpha)$ around the
point $(U^{[k]},\alpha^{[k]})$ and $J(U^{[k]},\alpha^{[k]})$ denotes
the Jacobian of $\sigma_{c\Delta T
}(\mathcal{F}(U^{[k]},\alpha^{[k]}))$.  Since $\mathcal{F}$ is
defined as a sequence of computational steps, it is impossible to
construct the Jacobian $J$ explicitly.  We therefore use
a Krylov method (GMRES) that only requires its application to a
vector $v$, which can be estimated as 
\begin{multline}
(I-J(U,\Delta T))v\approx\\
 v-\frac{\sigma_{c\Delta T}\left(\mathcal{F}(U+\epsilon v,\Delta T)\right)-\sigma_{c\Delta T}\left(\mathcal{F}(U,\Delta T)\right)}{\epsilon}.
\end{multline}

Since the convergence rate of GMRES depends
sensitively on the spectral properties of the system matrix (\ref{eq:linear_system}),
we propose to precondition
the linear system (\ref{eq:linear_system}) with a rough macroscopic
model based on a PDE to speed up the convergence \cite{samaey}.
In section \ref{sec:Chapman-Enskog-expansion-of}, we derived an
approximate PDE model using a Chapman-Enskog expansion.
We define a time-stepper for this approximate model as 
$F(U(x,t),\Delta T)$, 
%%that brings
%%$U(x,t)=(\rho(x,t),E(x,t))$ from $t$ to $\Delta T$.  This time-stepper
%%is similar to the coarse-grained time-stepper (\ref{eq:cgts}).
%%Together with the shift-back operator 
we can again write a non-linear
system \begin{equation} G(U,\alpha)=\left\{
\begin{array}{lcc} U-\sigma_{c\Delta T}(F(U,\Delta T)) & = & 0\\ p(U)
& = & 0\end{array}\right., \label{eq:non-linear system
preconditioner}\end{equation} 
in which we have replaced the coarse-grained time-stepper
by a time-stepper for the approximate PDE model. %%instead of
The solution of this system will
look very similar to the solution of the full model, but will differ
in places where the approximations made during the derivation of the
PDE model fail.
\begin{figure}
\includegraphics{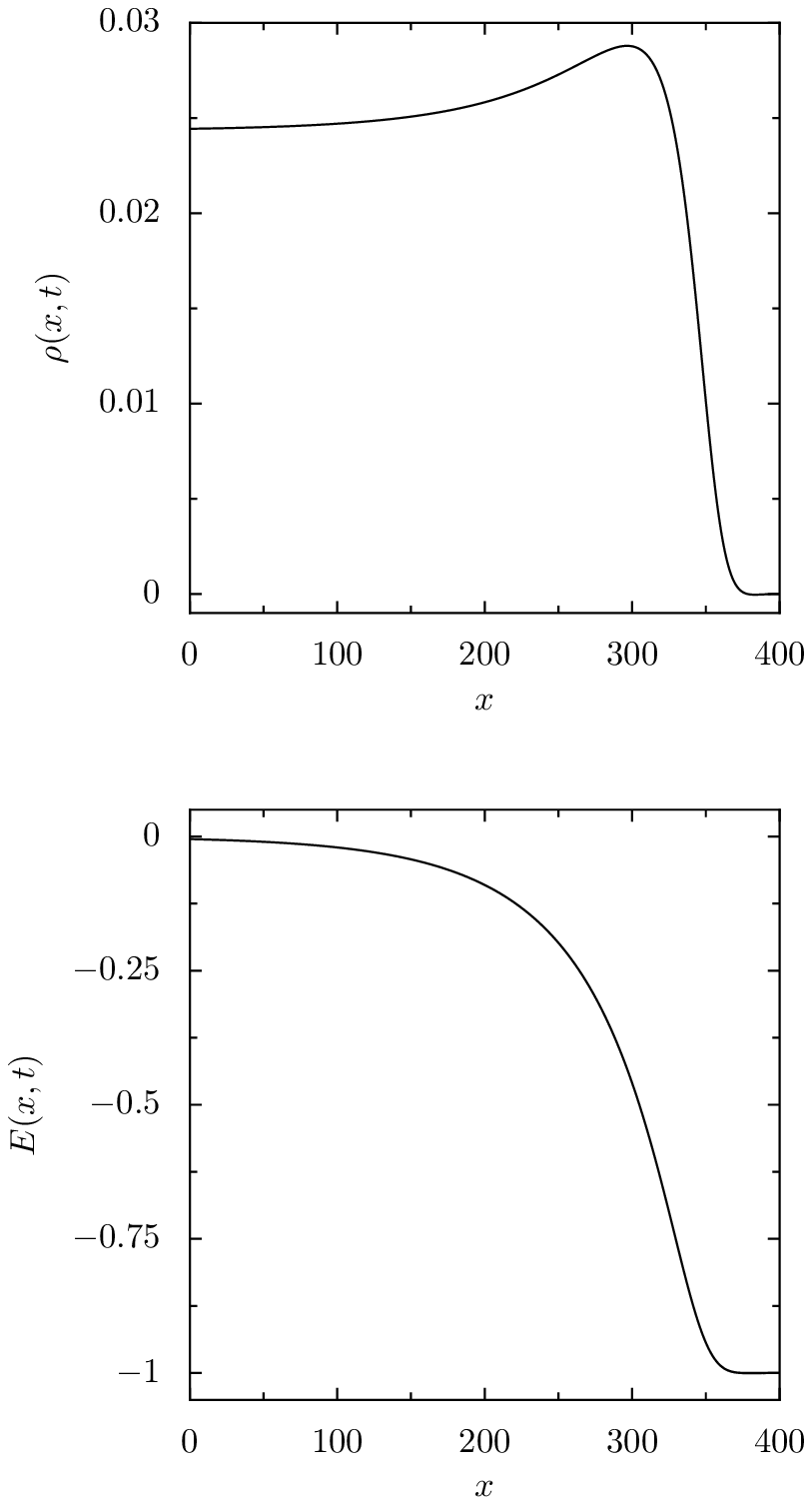}
\caption{\label{wave_micro}
The traveling wave solution for $c=1.3$ and $R=100$.  It is a fixed
point of sequential application of the evolution with the
coarse-grained time-stepper over a time $\Delta T$ and the shift-back
over a distance $c\Delta T$. }
\end{figure}
The linearization of this problem leads to a matrix problem for the
Newton corrections $dU^{[k]}$ and $d\alpha^{[k]}$ as in (\ref{eq:linear_system})
with very similar spectral properties. The Jacobian however, is now
known analytically and can be inverted easily. The idea is to
use this matrix as a preconditioner of the linear system that is solved
each Newton iteration. The preconditioned system reads
\begin{eqnarray}
\left( M(U^{[k]},\alpha^{[k]}) \right)^{-1} A(U^{[k]},\alpha^{[k]})\left(\begin{array}{c}
dU^{[k]}\\ d\alpha^{[k]}\end{array}\right)\nonumber \\
=-\left( M(U^{[k]},\alpha^{[k]}) \right)^{-1} G(U^{[k]},\alpha^{[k]}),
\end{eqnarray}
where $\left(M(U^{[k]},\alpha^{[k]})\right)^{-1}$ is the inverse of the matrix of
the linearization of (\ref{eq:non-linear system preconditioner}) and
$A(U^{[k]},\alpha^{[k]})$ denotes the linear system of
(\ref{eq:linear_system}).  Because the spectral properties of
$A(U^{[k]},\alpha^{[k]})$ and $M(U^{[k]},\alpha^{[k]})$ are so similar
that $\left(M(U^{[k]},\alpha^{[k]})\right)^{-1}A(U^{[k]},\alpha^{[k]})$
has spectral properties favorable for the convergence of GMRES.
Detailed numerical experiments, showing the spectral properties of the linear systems and the GMRES convergence, are reported in \cite{samaey}.

\subsection{The minimal speed and the coarse-grained time-stepper}
\label{sec:critical_speed}

%%In section \ref{sub:traveling-wave-solution}, we derived the minimal
In section IV, we derived the minimal
speed $c^{*}$ \eqref{minimal_speed} of the uniformly translating front solution
of the PDE model in terms of the asymptotic transport coefficients.
Solutions with a speed below this critical value have a density that
oscillates in the asymptotic region, which leads to unphysical negative
densities.

For the coarse-grained time-stepper, the asymptotic transport
coefficients are not available and no analytic expression for the
minimal speed can be found. We propose to use the time-stepper and its
fixed point solutions to determine the minimal speed. We vary the
imposed speed of the shift-back operator $\sigma_{c\Delta T}$ and
monitor the solutions of the fixed point problem in the asymptotic
region. As the imposed speed $c$ falls below the
minimal speed $c^{*}$ the solutions will become oscillatory because
the solution in the asymptotic region is a combination of two
exponentials \eqref{combexp},
%%$Ae^{\lambda_{+}x}+Be^{\lambda_{-}x}$ , 
whose exponents $\lambda_{\pm}$ coalesce and become complex at the
critical speed.

We can extract the two exponents from the solution in the asymptotic
region, if we assume that it fits a second order ODE of the form \[
\frac{\partial^{2}\rho}{\partial x^{2}}=a_{1}\rho+a_{2}\frac{\partial\rho}{\partial x}.\]
The coefficients $a_{1}$ and $a_{2}$ are found by taking the fixed point solution in two grid points, where we estimate the spatial derivatives using finite differences. This allows us to formulate a 2-by-2 system for $a_1$ and $a_2$.

The eigenvalues of the 2-by-2 matrix 
\begin{equation}
\label{lambd_minspeed}
\left(\begin{array}{cc}
0 & 1\\ a_{1} & a_{2}\end{array}\right)
\end{equation} 
are then $\lambda_{+}$ and $\lambda_{-}$, which coalesce at the critical speed.

For a speed $c$ far above the critical speed $c^{*}$, this method
estimates only one of the two exponents with confidence. Indeed, above
the critical speed the solution in the asymptotic region is a
combination of two decaying exponentials with different
exponents. However, one of them is slowly decaying, while the other
decays fast. Far away from the critical speed, the method will only
detect the slowly decaying exponential and a fit with a first order
ODE would be sufficient.  Near the critical speed, however, the
solution is a combination of both exponentials that decay with comparable
rates and both eigenvalues can be reliably extracted from the solution.

% ==========================================
%SUBSECTION:  FIXED POINT ALGORITM.
% ===========================================

\section{Numerical Results \label{sec:Numerical-Results}}

As an illustration, we look at a one-dimensional lattice Boltzmann
model on a grid with $N=1600$, grid distance
$\Delta x=0.4$ and a time step $\Delta
t=0.008$. We look at a model with five velocities with
($\mathcal{S}=\{-2,-1,0,-1,2\})$, weights
$w_{i}^{eq}=\{0,1/4,2/4,1/4,0\}$ and an electron diffusion coefficient
of $D=1.0$.  This leads to a relaxation parameter of $\tau=0.8$ or
$\omega=1.25$. Note that with this choice of equilibrium weights only
slow particles exist in the absence of external fields.

We enforce boundary conditions at the level of the lattice Boltzmann
model where we use homogeneous Dirichlet boundaries at the right and
no-flux boundaries at the left. The electrical field is kept constant
at $E^{+} = -1.0$ at the right and at the left we require that
$\partial^2 E/\partial x^2 = 0$.

We first compute the traveling wave with speed $c=1.30$ for a reaction
rate $R=100$, which is shown in figure \ref{wave_micro}.  As an initial guess
for the Newton procedure, we take 
\begin{equation}
\begin{array}{ccc}
	\rho(x,t) &=& 0.025/(1+\exp(0.15(x-L \, 2/3)))\\
	E(x,t)   &=& -1/(1+\exp(0.05(x-L \, 5/9)))
\end{array}
\end{equation}
To assess the overall performance of the method, we compute the total number of required lattice Boltzmann  time steps. 
We observe that about 5 Newton steps are needed. Each Newton step,
in turn, requires the solution of a linear system with the help of a
preconditioned Krylov subspace.  On average about 40 GMRES iterations
are required to solve the linear system with a tolerance of $1\cdot 10^{-12}$. 
Each GMRES iteration requires an evaluation of the
CGTS, which costs about 50 lattice Boltzmann iterations --- 25
for the lifting and 25 for the simulation.  This leads to a total of
10 000 evolutions with the lattice Boltzmann system to find a single
fixed point.  Detailed figures about convergence are given in
\cite{samaey}.

Next, we vary the cross section $R$, which is related to the number of
times an ionization reaction appears, and study the effect on the
critical velocity of the traveling wave solution. We  increase
$R$ from 60 to 100, which corresponds at the level of the PDE to an
increase of the growth transport coefficient from
$\alpha(E^+)=0.02015$ to $\alpha(E^+)=0.03707$, with $E^+=-1.0$. For
this range of reaction rates with the chosen value of $\Delta t$, 
the constrained runs algorithm always converges because all the eigenvalues of the Jacobian are 
smaller than $|1-1/\tau-\Delta tR|<1$, as discussed in
section~\ref{sec:lifting}.

We now turn to the numerical computation of the critical velocity with
the method proposed in section \ref{sec:critical_speed}. In table
\ref{critical-speed}, we show the critical speed determined for a
series of ionization strengths between $R=60$ and $R=100$. As outlined in section
\ref{sec:critical_speed}, the critical velocity for each value of $R$ is found by performing
a numerical continuation with the wave speed $c$ as a free parameter and monitoring
the eigenvalues of the 2-by-2 matrix (\ref{lambd_minspeed}).
For comparison, we also show the minimal speed as obtained using the approximate
analytical expressions for the transport coefficients, which we found through the Chapman-Enskog expansion.
We observe a small difference between the two results. The critical speed that results from
the Chapman-Enskog expansion is higher than the critical speed that is computed using the CGTS.
Note that both methods make approximations and it is not
clear which method is more exact. 

A short discussion about the accuracy of the shift-back operator is
necessary. Since we have implemented the shift-back operator using
a forward Euler approximation of the characteristic solutions of the equation $ u_t + c u_x =0$, the
error in the critical speed grows
linearly with the time step $\Delta T$.  Therefore, we should take $\Delta T$ as small as possible,
i.e.~ the minimal possible number of lattice Boltzmann steps in the
simulation step. (Note that accuracy and efficiency go together here.) 
However, we cannot take less than 25 LBM steps
because these steps are necessary to reduce the error in the lifting
(see figure \ref{figure_crl}).

%\begin{figure}
%\resizebox{8cm}{5cm}{\includegraphics{speed_crit}}
%\caption{
%The critical speed as a function of the velocity dependent ionization
%rate $R$.  The solid line is the prediction of the Chapman-Enskog
%expansion, the points are the prediction with the coarse-grained time-stepper. 
%Note that below $R=60$ the ionization reaction becomes weaker
%and the front of the traveling wave becomes broader and broader as
%diffusion dominates. The front does not fit anymore on the choose length of the grid.  \label{critical-speed}}
%\end{figure}
\begin{table}
\begin{tabular}{c|cc}
$R$ &  $c^*$ with   & $c^*$ with\\
& Chapman-Enskog & CGTS\\
\hline 
60   &1.3351 & 1.3177  \\
70   &1.3496 & 1.3318\\
80   &1.3609 & 1.3433\\
90   &1.3696 &  1.3517\\
100 &1.3755 &1.3566
\end{tabular}
\caption{
The critical speed for different values of the ionization
rate $R$. \label{critical-speed}}
\end{table}

%

%\begin{figure}
%\resizebox{8cm}{8cm}{\includegraphics{electric_field_vs_alpha}}
%\caption{The transport coefficient through. This needs to be compared with the exponential of the thownsend approximation}
%\end{figure}

%dummy comment inserted by tex2lyx to ensure that this paragraph is not empty
%dummy comment inserted by tex2lyx to ensure that this paragraph is not empty
%dummy comment inserted by tex2lyx to ensure that this paragraph is not empty

\section{Discussion and Conclusions}

In one dimension, an initial seed of electrons in a strong electrical
field will evolve into a streamer front that travels with a constant
speed $c$. Before the front, the density is zero and the electrical
field is constant; behind the front the electrical field is shielded and
there is a surplus of electrons.
In this article we extended the minimal streamer model of Ebert et
al.~\cite{ebert} to add some details about the microscopic
physics. To this end, we replaced the reaction-diffusion PDE with a lattice
Boltzmann model that has a velocity dependent reaction term.  The
reaction rates are a chosen to model the ionization reaction, where
fast particles have a given probability to undergo an ionization
reaction and create two slow particles. The
electrical field changes simultaneously as a result of the charge creation.

This macroscopic behavior of the model was analyzed with two methods.
The first is the more traditional analysis based on a
Chapman-Enskog expansion that derives an approximate PDE model and the
corresponding transport coefficients.  The resulting PDE model is very
similar to the minimal streamer model of Ebert, Van Saarloos and
Caroli, where the ionization rate depends on the local electrical
field. Based on the transport coefficients, we found the dependence of
the critical
speed (below which the traveling waves are unphysical)
and its dependence on the strength of the ionization cross
section.

Our second method is a computational method based on the
coarse-grained time-stepper, which defines the effective evolution law as a
sequence of computational steps.  The traveling wave solutions were
formulated as fixed points of this coarse-grained time-stepper
combined with a shift-back operator. By varying the applied shift-back, we again
found the critical speed, and we demonstrated how to calculate its dependence on
the cross section strength numerically. 

We showed that the coarse-grained time-stepper provides a viable alternative
to the traditional Chapman-Enskog analysis.  However, in this paper, 
we did not make any statements about which of the
two methods provides the most accurate information.  Both methods make
approximations to obtain the critical velocity of the proposed 
lattice Boltzmann model.  In contrast to the theoretical approximations
in the Chapman-Enskog analysis, these approximations are due to numerical accuracy
for the coarse-grained time-stepper.  

Further progress in the accuracy can be made in several ways. First,
the lifting can be done more accurately if a higher order constrained
runs algorithm is used. Such a method would take multiple steps with the
lattice Boltzmann model and uses these multiple points to provide us
with an approximate initial state \cite{gear}. We expect that these
higher order lifting procedures will reproduce more than two terms in
the Chapman-Enskog series and therefore provide a better initial state for the
simulation step. This will allow limiting the number of simulation
steps, which will immediately improve the accuracy of the computed solutions and the critical speed.
%%A second way to improve the accuracy is to implement the shift-back operator with, for
%%example, the help of a Fourier transform. A third possible increase in
%%accuracy can be obtained with a better extraction of the eigenvalues
%%of the asymptotic solution. The improvement of the accuracy and
%%performance is the focus of our future research.

Although the model problem in this paper is non-trivial, and the Chapman-Enskog expansion is tedious,
we emphasize that our method is mainly developed with applications in sight where the
traditional Chapman-Enskog analysis is completely intractable.  The model problem in this paper
both allows us to analyze the proposed methods and provide directions for further development.
In a forthcoming article, we will do a complete analysis of the traveling
wave solutions in the proposed model and illustrate more extensively how the macroscopic
behavior depends on the parameters of the microscopic model. In our
calculations, we have also observed the positively charged fronts
that move in the opposite direction of the electrical field. These
solutions have been discussed by Ebert et al in \cite{ebert} and
our methods are also able to locate these states. 

We expect that most of the results presented in this paper will
remain valid when the number of velocities in the discretization of
the Boltzmann model is increased.  With additional velocities, it is
possible to model the cross sections of colliding
particles with more detail and study other reactions than the ionization reaction. It
would also be interesting to study models with multiple species where
the collision rates between the particles are velocity dependent.
This would allow us to include photo-ionization effects into the minimal
streamer model, an important effect that is neglected in the current model 
\cite{kuli}.

\begin{acknowledgments}
GS is a postdoctoral researcher of the Fund for Scientific Research ---
Flanders, who also supports PVL through projects G.0130.03 and
G.0365.06. The paper presents research results of the Belgian
Programme on Interuniversity Attraction Poles, initiated by the
Belgian Federal Science Policy Office, which also funded WV with a
DWTC return grant. We also thank Christophe Vandekerckhove for fruitful discussions.
\end{acknowledgments}

\end{document}